\documentclass[prd,preprint,tightenlines,floatfix,
preprintnumbers,nofootinbib,eqsecnum,superscriptaddress]{revtex4}

\usepackage[T1]{fontenc}		% fnot encoding with polish letters !or OT4

\usepackage{amsmath,amsfonts,amssymb,amstext,mathrsfs}
\usepackage{mathpazo}

\usepackage[dvips]{graphicx}
\usepackage{epsf,float}
\usepackage{revsymb}

\usepackage{dcolumn}% Align table columns on decimal point
\usepackage{braket}
\usepackage{color,xcolor}
\usepackage{graphicx}
\usepackage{subfigure}
\usepackage{multirow}
\usepackage{tabularx}
\usepackage{pstricks}
\usepackage[section]{placeins}
\usepackage{booktabs}
\usepackage{array}

\usepackage{hyperref}
\newcommand{\Pom}{\mathbb{P}}

\renewcommand\slash[1]{\not \! #1}

\newcommand{\bp}{\mbox{\boldmath $p$}}
\newcommand{\bk}{\mbox{\boldmath $k$}}
\newcommand{\bq}{\mbox{\boldmath $q$}}

\begin{document}

\title{
Exclusive and semiexclusive production of \boldmath{$\mu^+ \mu^-$} pairs\\
with \boldmath{$\Delta$} isobars and other resonances in the final state
\\
and the size of absorption effects}

\author{Piotr Lebiedowicz}
 \email{Piotr.Lebiedowicz@ifj.edu.pl}
\affiliation{Institute of Nuclear Physics Polish Academy of Sciences, 
ul. Radzikowskiego 152, PL-31-342 Krak\'ow, Poland}

\author{Antoni Szczurek
\footnote{Also at \textit{Faculty of Mathematics and Natural Sciences, 
University of Rzesz\'ow, ul. Pigonia 1, PL-35-310 Rzesz\'ow, Poland}.}}
\email{Antoni.Szczurek@ifj.edu.pl}
\affiliation{Institute of Nuclear Physics Polish Academy of Sciences, 
ul. Radzikowskiego 152, PL-31-342 Krak\'ow, Poland}

\begin{abstract}
We include $p p \to p \Delta^{+} \mu^{+} \mu^{-}$ and
$p p \to \Delta^{+} \Delta^{+} \mu^{+} \mu^{-}$ processes
in addition to the standard $p p \to p p \mu^{+} \mu^{-}$ process
both in equivalent-photon approximation (EPA) 
and in exact $2 \to 4$ calculations. 
For comparison we calculate also the continuum proton dissociation 
in a recently developed $k_{t}$-factorization
approach to $\gamma \gamma$-involved processes with parametrizations 
of $F_{2}$ structure function known from the literature.
The calculated cross section including all the processes
is considerably larger than the one measured recently by 
the ATLAS collaboration. 
We calculate absorption effects for $p p \to p p \mu^{+} \mu^{-}$ process 
in the momentum space.
The final cross section with absorption effects is by 10\% larger 
than the one measured by the ATLAS collaboration 
which is difficult to explain.
Several differential distributions with the ATLAS
experimental cuts are presented.
It is shown that the processes with electromagnetic 
$p \to \Delta(1232)$ and $p \to N(1440)$ transitions, that have similar
characteristics as the $p p \to p p \mu^+ \mu^-$ process,
increase the cross section for $\mu^+ \mu^-$ production
and thus can affect its theoretical interpretation
when the leading baryons are not detected as is the case for 
the CMS and ATLAS measurements.
The mechanism of dissociation into hadronic continuum 
is not under full control as the corresponding absorption effects
are not fully understood.
We present first predictions for future ATLAS experiment 
with the ALFA sub-detectors and discuss observables relevant 
for this measurement.
\end{abstract}

\maketitle

%----------------------------
\section{Introduction}
\label{sec:intro}
%----------------------------

Recently the ATLAS collaboration measured production of
muon pairs with the requirement of rapidity gap 
and small transverse momentum of the dimuon system
at proton-proton collision energy $\sqrt{s} = 13$~TeV~\cite{Aaboud:2017oiq}. 
A similar study was done previously at $\sqrt{s} = 7$~TeV 
\cite{Chatrchyan:2011ci,Chatrchyan:2012tv,Aad:2015bwa}.
There have been also efforts to install and use forward proton
detectors, see e.g. \cite{Cms:2018het}.
In order to ensure the exclusivity of dimuon measurement \cite{Aaboud:2017oiq}
%curious procedure have been imposed by ATLAS collaboration \cite{Aaboud:2017oiq},
%among other things, 
the measurement was performed for a dimuon invariant mass
of $12 \;{\rm GeV} < M_{\mu^+ \mu^-} < 70\;{\rm GeV}$ with different
$p_{t,\mu}$ conditions and the muon pair was required to have 
a transverse momentum $p_{t,\mu^+ \mu^-} <1.5$~GeV.
It is believed that such requirements cause that the cross section is dominated by 
the $p p \to p p \mu^+ \mu^-$ fully exclusive contribution.

The common approach to calculate cross sections for photon induced processes is
the equivalent-photon approximation (EPA).
The ATLAS collaboration observed significantly lower cross section
than that predicted by the EPA approach \cite{Aaboud:2017oiq}.
The effect could be caused by absorption effects which destroy rapidity gaps. 
The absorption effects can be calculated in the momentum space
(see e.g. \cite{Lebiedowicz:2012gg,Lebiedowicz:2015cea,Lebiedowicz:2016lmn}) 
or in the impact parameter space (see e.g. \cite{Dyndal:2014yea}).
Only very few differential distributions can be obtained in the EPA approach.
In the impact parameter space approach the situation is similar \cite{Dyndal:2014yea}.
However, experimental cuts on (pseudo)rapidities and
transverse momenta of muons selected only some kinematical configurations, 
so the use of the phase space averaged value of the gap survival factor may be not justified.
Moreover, the effect of absorption strongly depends on kinematics of outgoing protons
\cite{Schafer:2007mm,Lebiedowicz:2015cea}.
Therefore in the following we perform precise calculation
of the exclusive $2 \to 4$ process (8-fold phase space integration) 
as is routinely done for instance for the $p p \to p p \pi^+ \pi^-$ reaction
\cite{Lebiedowicz:2009pj,Lebiedowicz:2015eka,Lebiedowicz:2016ioh},
for the $p p \to p p K^+ K^-$ reaction \cite{Lebiedowicz:2011tp,Lebiedowicz:2018eui}
or for the $p p \to p p p \bar{p}$ reaction \cite{Lebiedowicz:2018sdt}.

The exclusive $p p \to p p \mu^+ \mu^-$ process competes with the two-photon interactions
involving single- and double-proton dissociation contributions.
The ATLAS experiment imposes a rather loose cut on $p_{t,\mu^+ \mu^-} < 1.5$~GeV \cite{Aaboud:2017oiq}. 
To which extend such a loose cut allows participation of other
mechanisms such a single or double proton resonance production? 
The inclusion of $\Delta$ isobar seems potentially the most important.
The electromagnetic production of one $\Delta$ resonances on either leg
or simultaneous excitation of two $\Delta$ resonances on both legs was never 
discussed quantitatively in the context of semiexclusive production of dilepton pairs.
EPA fluxes of photons associated with $\Delta$ production were presented in \cite{Baur:1998ay}. 
They were used only in \cite{Guzey:2014axa} for the $p p \to p \Delta J/\psi$ process. 
Semiexclusive processes with the $\Delta$ excitations for 
the $J/\psi$ production were estimated also in \cite{Cisek:2016kvr}.

During the last years several authors have discussed 
the backgrounds for the process $p p \to p p (\gamma \gamma \to \mu^+ \mu^-)$.
In particular, the contribution of the semielastic and
inelastic $\mu^+ \mu^-$ production, where one or both protons dissociate, 
have been analyzed in \cite{daSilveira:2014jla,Luszczak:2015aoa,daSilveira:2015hha,Goncalves:2018gca}.
The proton-dissociative processes have significantly different 
kinematic distributions compared to the elastic (purely exclusive) process,
which allows in principle for a separation of the different production mechanisms.

In the present paper we wish to consider
the $p p \to p \Delta \mu^+ \mu^-$ and $p p \to \Delta \Delta \mu^+ \mu^-$ processes 
both in EPA and in exact $2 \to 4$ calculation in addition
to the standard $p p \to p p \mu^+ \mu^-$ process.
For reference we shall include also production of Roper resonance ($N(1440)$) 
for which some knowledge is available from the studies with the CLAS detector
at the Thomas Jefferson National Accelerator Facility (JLab) \cite{Aznauryan:2009mx,Mokeev:2015lda}.
In the following we wish to discuss absorption effects 
for the $2 \to 4$ processes considered.
We wish to estimate also the role of electromagnetically induced
proton-dissociative processes ($p p \to p X \mu^+ \mu^-$ and $p p \to XY \mu^+ \mu^-$)
calculated in a recently developed $k_t$-factorization approach using the
phenomenological parametrizations of the deep-inelastic structure functions 
from the literature.

%----------------------------------------
\section{Theoretical approach}
\label{sec:approach}
%----------------------------------------

%------------------------------------------------------------------
\begin{figure}
(a)\includegraphics[width=5.cm]{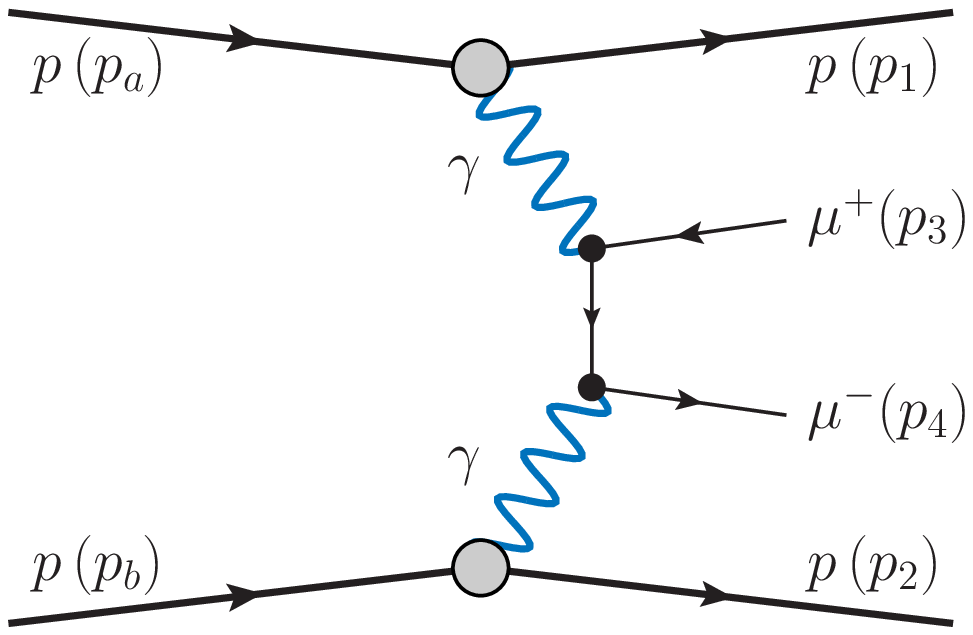}  
(b)\includegraphics[width=5.cm]{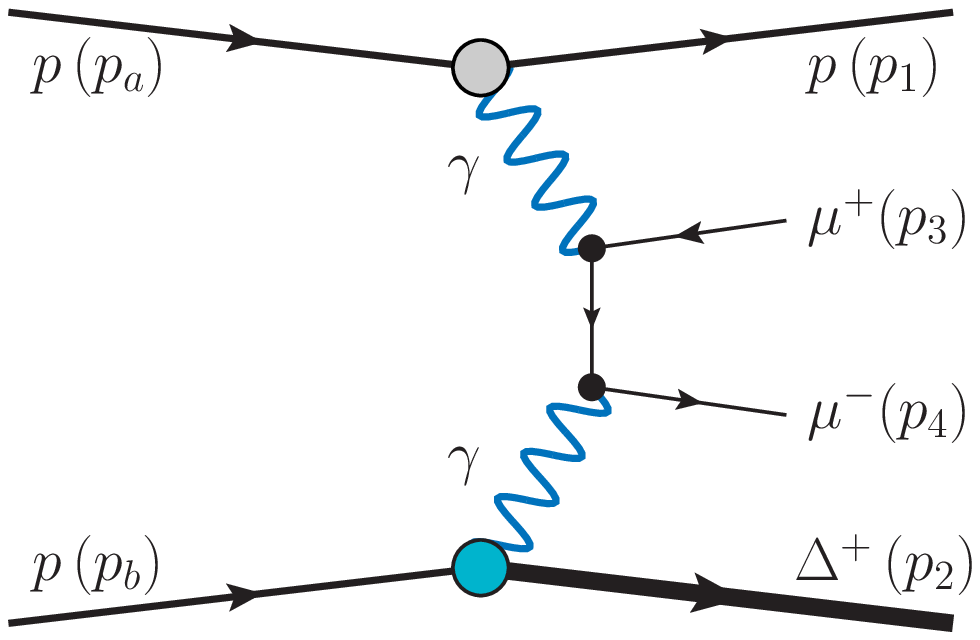}\\
(c)\includegraphics[width=5.cm]{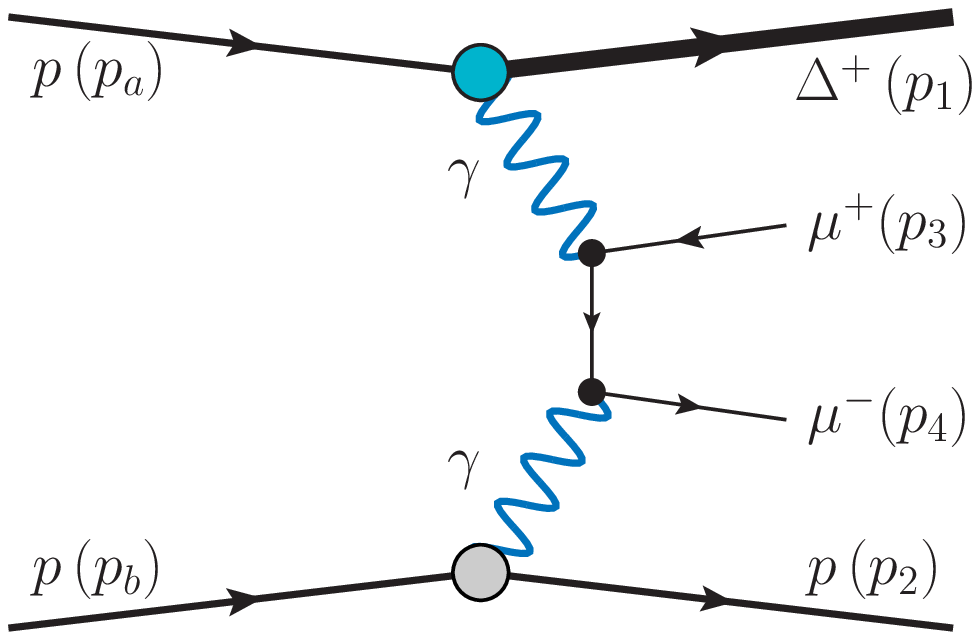}
(d)\includegraphics[width=5.cm]{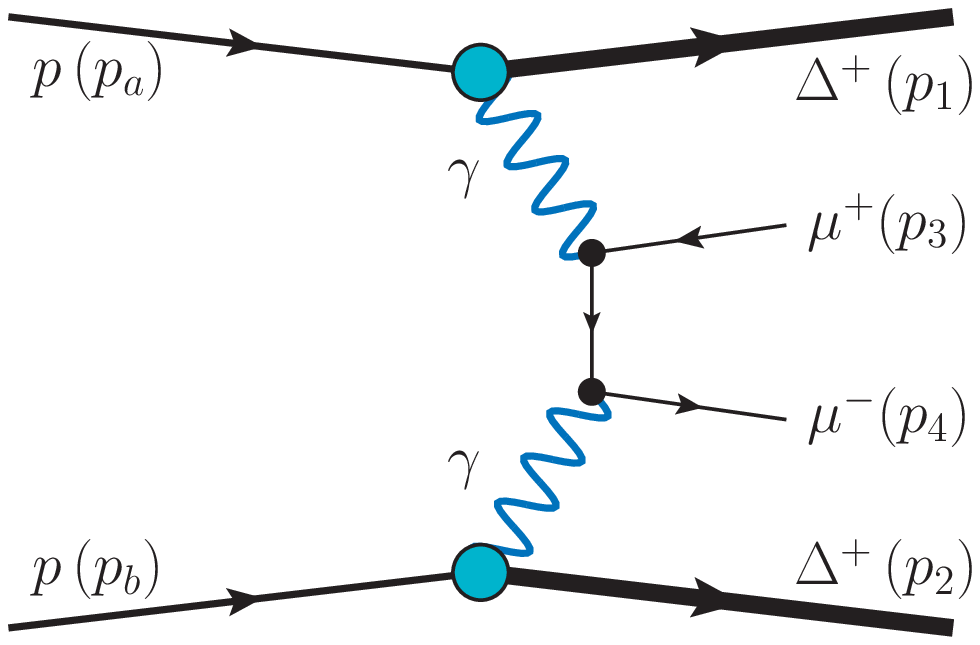}    
  \caption{\label{fig:diagrams}
  \small 
Diagrams for selected exclusive processes 
for two-photon production of muon pairs in $pp$ collisions on the Born level.
Here only the $\hat{t}$-channel diagrams are shown. 
There are also corresponding $\hat{u}$-channel diagrams with the
photon-muon vertices interchanged.
}
\end{figure}
%-------------------------------------------------------------------

In Fig.~\ref{fig:diagrams} we show ``Born level'' diagrams 
of processes considered in the present analysis for central exclusive 
$\mu^+ \mu^-$ production in proton-proton collisions:
\begin{eqnarray}
&&p + p \to p + \mu^{+} + \mu^{-} + p\,,
\label{2to4_reaction_pp}\\
&&p + p \to p + \mu^{+} + \mu^{-} + \Delta^{+}\,,
\label{2to4_reaction_pD}\\
&&p + p \to \Delta^{+} + \mu^{+} + \mu^{-} + p\,,
\label{2to4_reaction_Dp}\\
&&p + p \to \Delta^{+} + \mu^{+} + \mu^{-} + \Delta^{+}\,.
\label{2to4_reaction_DD}
\end{eqnarray}
Only the process (\ref{2to4_reaction_pp}) shown by the diagram (a)
was considered so far in the literature.

In the following we will calculate the contributions from the diagrams
shown in Fig.~\ref{fig:diagrams}.

%----------------------------------------
\subsection{Exact $2 \to 4$ kinematics}
\label{exact}
%----------------------------------------

In the present studies we perform, for the first time, exact 
calculations for all the considered exclusive $2 \to 4$ processes
shown in Fig.~\ref{fig:diagrams}.
In general, the cross section can be written as
\begin{eqnarray}
d \sigma &=& 
\frac{(2 \pi)^{4}}{2s} 
{\overline{|{\cal M}_{2 \to 4}|^2}}
\frac{d^3 p_1}{(2 \pi)^{3} 2 E_1} 
\frac{d^3 p_2}{(2 \pi)^{3} 2 E_2}
\frac{d^3 p_3}{(2 \pi)^{3} 2 E_3} 
\frac{d^3 p_4}{(2 \pi)^{3} 2 E_4} 
\nonumber \\
&&\times \delta^{4} \left(E_a + E_b -p_1 - p_2 - p_3 - p_4 \right)  \,.
\label{differential_cs}
\end{eqnarray}
The formula (\ref{differential_cs}) is written in the overall center-of-mass frame
where energy and momentum conservations have been made explicit, see \cite{Lebiedowicz:2009pj}.
The phase space integration variables are taken the same as in \cite{Lebiedowicz:2009pj}, 
except that proton transverse momenta $p_{t,1}$ and $p_{t,2}$ are replaced by
$\log_{10}(p_{t,1}/p_{t,0})$ and $\log_{10}(p_{t,2}/p_{t,0})$, $p_{t,0} = 1$~GeV, respectively.
%The main ingredients of the model are the amplitudes for the exclusive process. 
In (\ref{differential_cs}) $\overline{|{\cal M}_{2 \to 4}|^2}$ is the $2 \to 4$ amplitude squared 
averaged over initial and summed over final particle polarization states.
The kinematic variables for the $2 \to 4$ reaction are
\begin{eqnarray}
&&s = (p_{a} + p_{b})^{2}\,, 
\quad s_{34} = M_{\mu^{+}\mu^{-}}^{2} = (p_{3} + p_{4})^{2},\,\nonumber \\
&& q_1 = p_{a} - p_{1}\,,  \quad q_2 = p_{b} - p_{2}\,, 
\quad t_1 = q_{1}^{2}\,,  \quad t_2 = q_{2}^{2}\,;
\label{2to4_kinematic}\\
&&
\hat{p}_{t} = p_{4} - q_{2} = q_{1} - p_{3}\,, 
\quad \hat{p}_{u} = q_{2} - p_{3} = p_{4} - q_{1}\,, \quad
\hat{t} = \hat{p}_{t}^{2}\,, \quad \hat{u} = \hat{p}_{u}^{2}\,,
\label{2to4_kinematic_tu}
\end{eqnarray}
where $p_{a,b}$ and $p_{1,2}$ denote the four-momenta of the baryons, 
and $p_{3,4}$ denote the four-momenta of the muons, respectively.

The Born amplitudes for the processes 
(\ref{2to4_reaction_pp}) - (\ref{2to4_reaction_DD}) are calculated as
\begin{eqnarray}
{\cal M}_{\lambda_a \lambda_b \to \lambda_1 \lambda_2 \lambda_3 \lambda_4}(t_{1},t_{2}) =
V_{\lambda_a \to \lambda_1}^{\mu_1}(t_{1})\;
\frac{g_{\mu_1 \nu_1}}{t_1} \;
V_{\lambda_3 \lambda_4}^{\nu_1 \nu_2}\;
\frac{g_{\nu_2 \mu_2}}{t_2}\;
V_{\lambda_b \to \lambda_2}^{\mu_2}(t_{2}) \,,
\label{born}
\end{eqnarray}
where $\lambda_{a,b}$, $\lambda_{1,2} = \pm \frac{1}{2}$ 
denote the helicities of the baryons, 
and $\lambda_{3,4} = \pm \frac{1}{2}$ denote the helicities of the muons, respectively.
The $\gamma \gamma \to \mu^+ \mu^-$ interaction
includes both $\hat{t}$- and $\hat{u}$-channel amplitudes:
\begin{eqnarray}
V_{\lambda_3 \lambda_4}^{\nu_1 \nu_2} = -e^{2} \, \bar{u}(p_{4},\lambda_{4}) 
\left( 
  \gamma^{\nu_{2}} \frac{\slash{\hat{p}_{t}} + m_{\mu}}{\hat{t} - m_{\mu}^{2}} \gamma^{\nu_{1}}
+ \gamma^{\nu_{1}} \frac{\slash{\hat{p}_{u}} + m_{\mu}}{\hat{u} - m_{\mu}^{2}} \gamma^{\nu_{2}}
\right) v(p_{3},\lambda_{3}) \,,
\label{amp_tu}
\end{eqnarray}
where $\bar{u}(p_{4},\lambda_{4})$ and $v(p_{3},\lambda_{3})$ 
are muon ($\mu^{-}$) and antimuon ($\mu^{+}$) spinors, respectively.

The $\gamma pp$ vertex is written as
\begin{eqnarray}
V_{\lambda \to \lambda'}^{(\gamma pp) \mu}(p',p) = e \, \bar{u}(p',\lambda') 
\left( 
\gamma^{\mu} F_{1}(t) + \frac{i \sigma^{\mu \nu} (p'-p)_{\nu}}{2 m_{p}} F_{2}(t)
\right) u(p,\lambda) \,,
\label{vertex_spinors}
\end{eqnarray}
where $u(p,\lambda)$ is a Dirac spinor and $p, \lambda$ and $p', \lambda'$ 
are initial and final four-momenta and helicities of the protons, respectively.
The form factors $F_{1}(t)$ and $F_{2}(t)$ correspond to the proton
helicity-conserving and helicity-flipping transitions.

The electromagnetic transition between a proton and spin 1/2 positive parity nucleon
resonance $N^{*}$, using the Dirac $F_{1}^{*}$ and Pauli $F_{2}^{*}$
type form factors, satisfying manifestly electromagnetic
gauge-invariance, can be written as \cite{Tiator:2008kd}:
\begin{equation}
\begin{split}
&V_{\lambda \to \lambda'}^{(\gamma pN^{*}) \mu}(p',p) = \\
&\qquad e \, \bar{u}^{(N^{*})}(p',\lambda') 
\left[ 
\left( \gamma^{\mu} -\frac{(\slash{p'}-\slash{p}) (p'-p)^{\mu}}{t} \right) F_{1}^{*}(t)
+ \frac{i \sigma^{\mu \nu} (p'-p)_{\nu}}{m_{N^{*}} + m_{p}} F_{2}^{*}(t)
\right] u(p,\lambda)\,,
\end{split}
\label{vertex_spinors_Roper}
\end{equation}
where $u^{(N^{*})}$ is the $N^{*}$ Dirac spinor.
The Dirac-type proton-Roper transition form factor $F_{1}^{*}(t)$ vanishes at $t = 0$ and stays positive at large $|t|$.
On the other hand, the Pauli-type form factor $F_{2}^{*}(0) \simeq -0.6$
and changes sign around $-t \simeq 1$~GeV$^{2}$.
%For simplicity we assume $F_{1}^{*}(t) \propto -t/(m_{N^{*}} + m_{p})^{2}\, G_{D}(t)$.
We take analytic parametrizations for the electromagnetic $p \to N^{*}(1440)$ transition form-factors 
from \cite{Ramalho:2017muv};
see also Refs.~\cite{Tiator:2008kd,Ramalho:2014hia,Gutsche:2017lyu,Ramalho:2017pyc}.

The $\gamma p \Delta$ vertex can be written as \cite{Jones:1972ky}
\begin{eqnarray}
V_{\lambda \to \lambda'}^{(\gamma p \Delta) \mu}(p',p) 
= e \, \bar{u}_{\alpha}^{(\Delta)}(p',\lambda') \Gamma^{\alpha \mu} \,u(p,\lambda)\,,
\label{vertex_GND1}
\end{eqnarray}
where
\begin{eqnarray}
\Gamma^{\alpha \mu} = G_{M}^{*}(q^{2}) K_{M}^{\alpha \mu} + G_{E}^{*}(q^{2}) K_{E}^{\alpha \mu} + G_{C}^{*}(q^{2}) K_{C}^{\alpha \mu}\,,
%&& K_{M}^{\alpha \mu} = -\frac{3 (m_{\Delta} + m_{p})}{2 m_{p} \left[ (m_{\Delta} + m_{p})^{2} - q^{2} \right]}
%\varepsilon^{\alpha \mu \rho \sigma} \,P_{\rho} \,q_{\sigma}\,,
%&& \;\;\, \qquad \qquad + G_{E}(q^{2}) \left(q^{\alpha} P^{\mu} - q \cdot P \, g^{\alpha \mu} \right) i \gamma_{5}\\
%&& \;\;\, \qquad \qquad + G_{C}(q^{2}) \left(q^{\alpha} q^{\mu} - q^{2} \, g^{\alpha \mu} \right) i \gamma_{5} \rbrace
\label{vertex_GND2}
\end{eqnarray}
in terms of the magnetic dipole $G_{M}^{*}$, electric quadrupole $G_{E}^{*}$,
and Coulomb quadrupole $G_{C}^{*}$ transition form factors. 
In (\ref{vertex_GND1}) $u_{\alpha}^{(\Delta)}$ denotes the
Rarita-Schwinger spinor of the spin-3/2 $\Delta$ isobar.
The $\gamma p \to \Delta^{+}$ transition is dominated by the magnetic 
dipole form factor.
%Analysis of photoproduction and electroproduction shows that 
%the electric and Coulomb quadrupole terms are negligible.
We consider therefore only the magnetic transition term with
\footnote{Note that in \cite{Jones:1972ky} authors define $q = p' - p$.
We have made allowance for this in writing Eq. (\ref{vertex_GND3}).}
\begin{eqnarray}
K_{M}^{\alpha \mu} = \frac{3 (m_{\Delta} + m_{p})}{4 m_{p} \left[ (m_{\Delta} + m_{p})^{2} - q^{2} \right]}
\epsilon^{\alpha \mu \rho \sigma} \,(p'+p)_{\rho} \,q_{\sigma}\,.
%&& \;\;\, \qquad \qquad + G_{E}(q^{2}) \left(q^{\alpha} P^{\mu} - q \cdot P \, g^{\alpha \mu} \right) i \gamma_{5}\\
%&& \;\;\, \qquad \qquad + G_{C}(q^{2}) \left(q^{\alpha} q^{\mu} - q^{2} \, g^{\alpha \mu} \right) i \gamma_{5} \rbrace
\label{vertex_GND3}
\end{eqnarray}
For the magnetic transition form factor we use the phenomenological
parametrization of Ref.~\cite{Gail:2005gz}
\begin{eqnarray}
G_{M}^{*}(q^{2}) = 3 \, G_{D}(q^{2})\, \exp( 0.21 \,q^{2} ) \sqrt{1 - \frac{q^{2}}{(m_{\Delta} + m_{p})^{2}}}\,,
\label{vertex_GND4}
\end{eqnarray}
with the standard dipole form factor 
$G_{D}(t) = \left(1 - t/m_{D}^{2} \right)^{-2}$, $m_{D}^{2} = 0.71$~GeV$^{2}$.
The quality of the parametrization (\ref{vertex_GND4}) was studied in \cite{Ramalho:2008dp}.
%with static ($q^{2} = 0$) transition moment $G_{M}^{*}(0) = 3$.

%----------------------------------------------
\subsection{Absorption corrections}
\label{absorption}
%----------------------------------------------

The absorptive corrections to the Born amplitude (\ref{born})
are added to give the full physical amplitude for the $pp \to pp \mu^{+} \mu^{-}$ reaction: 
\begin{eqnarray}
{\cal {M}}_{pp \to pp \mu^{+} \mu^{-}} =
{\cal {M}}_{pp \to pp \mu^{+} \mu^{-}}^{{\rm Born}} + 
{\cal {M}}_{pp \to pp \mu^{+} \mu^{-}}^{{\rm absorption}}\,.
\label{amp_full}
\end{eqnarray}
Here (and above) we have for simplicity omitted 
the dependence of the amplitude on kinematic variables.

The amplitude including $pp$-rescattering corrections 
between the initial- and final-state protons within the one-channel eikonal approach 
can be written as
\begin{eqnarray}
{\cal M}_{pp \to pp \mu^{+} \mu^{-}}^{\mathrm{absorption}}(s,\bp_{t,1},\bp_{t,2})&=&
\frac{i}{8 \pi^{2} s} \int d^{2} \bk_{t} \,
{\cal M}_{pp \to pp}(s,-{\bk}_{t}^{2})\,
{\cal M}_{pp \to pp \mu^{+} \mu^{-}}^{\mathrm{Born}}(s,\tilde{\bp}_{t,1},\tilde{\bp}_{t,2})\,, \nonumber \\
\label{abs_correction}
\end{eqnarray}
where $\tilde{\bp}_{t,1} = {\bp}_{t,1} - {\bk}_{t}$ and
$\tilde{\bp}_{t,2} = {\bp}_{t,2} + {\bk}_{t}$.
Here, in the overall center-of-mass system, ${\bp}_{t,1}$ and ${\bp}_{t,2}$
are the transverse components of the momenta of the final-state protons
and ${\bk}_{t}$ is the transverse momentum carried by additional pomeron exchange.
${\cal M}_{pp \to pp}(s,-{\bk}_{t}^{2})$ 
is the elastic $pp$-scattering amplitude
for large $s$ and with the momentum transfer $t=-{\bk}_{t}^{2}$.
We assume $s$-channel helicity conservation in the pomeron-proton vertices.

In the following we shall show results in the Born approximation
as well as when including the absorption corrections on the amplitude level.
This allows us to study the absorption effects differentially in any
kinematical variable chosen for two-photon induced processes.

%---------------------------------------------------------------------
\subsection{Equivalent-photon approximation}
\label{EPA}
%---------------------------------------------------------------------

In the collinear EPA approach, with neglected photon transverse momenta,
one can write the differential cross section as
\begin{equation}
\frac{d \sigma}{d {\rm y}_3 \,d {\rm y}_4 \,d^{2}\bp_{t,\mu}} = \frac{1}{16 \pi^2 {\hat s}^2} 
x_1  f(x_1) x_2 f(x_2) \overline{|{\cal M}_{\gamma \gamma \to \mu^+ \mu^-}|^2} \,,
\label{EPA_formula}
\end{equation}
where ${\hat s} = s x_{1} x_{2}$. $f(x)$'s are elastic fluxes of 
the equivalent photons 
as a function of longitudinal momentum fraction with respect to the parent proton 
defined by the kinematical variables of the muons,
\begin{eqnarray}
&&x_1 = \frac{m_{t, 3}}{\sqrt{s}}\exp({\rm y}_3) + 
        \frac{m_{t, 4}}{\sqrt{s}}\exp({\rm y}_4) \,, \nonumber \\
&&x_2 = \frac{m_{t, 3}}{\sqrt{s}}\exp(-{\rm y}_3) + 
        \frac{m_{t, 4}}{\sqrt{s}}\exp(-{\rm y}_4) \,,
\label{fractions}
\end{eqnarray}
where $m_{t, \mu} = \sqrt{|\bp_{t,\mu}|^{2}+m_{\mu}^{2}}$.
In (\ref{EPA_formula}) $\overline{|{\cal M}|^2}$
is the $\gamma \gamma \to  \mu^+ \mu^-$ amplitude squared averaged over 
the photon and summed over the muon polarization states.
For the elastic photon fluxes $f(x)$ we take the formulas given in \cite{Budnev:1974de},
see \cite{Lebiedowicz:2015cea}.
The photon fluxes associated with $\Delta$ production are taken from Ref.~\cite{Baur:1998ay}.

%-----------------------------------------------------
\subsection{$k_t$-factorization approach}
\label{kt_approach}
%-----------------------------------------------------

In the recent $k_t$-factorization approach \cite{daSilveira:2014jla,Luszczak:2015aoa} 
the differential cross section for the $pp \to X \mu^+ \mu^- Y$ reaction
($X$ and $Y$ represent the hadronic systems resulting from the proton dissociation)
can be written as
\begin{eqnarray}
\frac{d \sigma(pp \to X \mu^+ \mu^- Y)}
{d{\rm y}_3 \,d{\rm y}_4 \,d^{2}\bp_{t,3} \,d^{2}\bp_{t,4} \, dM_{X} \,dM_{Y}} 
&=& \frac{1}{16 \pi^2 {\hat s}^2}
\int \frac{d^2 \bq_{t,1}}{\pi \bq_{t,1}^2} \frac{d^2 \bq_{t,2}}{\pi \bq_{t,2}^2}
x_{1} \frac{d\gamma(x_1,\bq_{t,1},M_X)}{dM_X}
x_{2} \frac{d\gamma(x_2,\bq_{t,2},M_Y)}{dM_Y} \nonumber \\
&&\times  
\sum_{\lambda_{3} \lambda_{4}} |M(\lambda_{3}, \lambda_{4}; \bq_{t,1},\bq_{t,2})|^2 \, 
\delta^{(2)}(\bq_{t,1} + \bq_{t,2} - \bp_{t,3} - \bp_{t,4})\,. \nonumber \\
\label{kt_factorisation}
\end{eqnarray}
Here $\bq_{t,1}$ and $\bq_{t,2}$ are the transverse momentum vectors of virtual photons.
The inelastic photon fluxes, $\gamma(x_1,\bq_{t,1},M_X)$ and $\gamma(x_2,\bq_{t,2},M_Y)$,
are expressed in terms of deep-inelastic structure functions \cite{Luszczak:2015aoa} 
known from many experiments.
%The $M_{X,Y}$-dependent photon fluxes can be decomposed into fluxes
%corresponding to the relevant proton dissociation.
Different parametrizations were proposed in the literature 
(see e.g. \cite{Luszczak:2015aoa} and references therein).
%It was shown in \cite{Luszczak:2015aoa} that
%different parametrizations give very different results.

%--------------------
\section{Results}
\label{sec:results}
%--------------------

%-------------------------------------------------------------------------
\subsection{Exact $2 \to 4$ approach} 
\label{sec:Exact_approach}
%-------------------------------------------------------------------------

\begin{figure}[!ht]
(a)\includegraphics[width=0.46\textwidth]{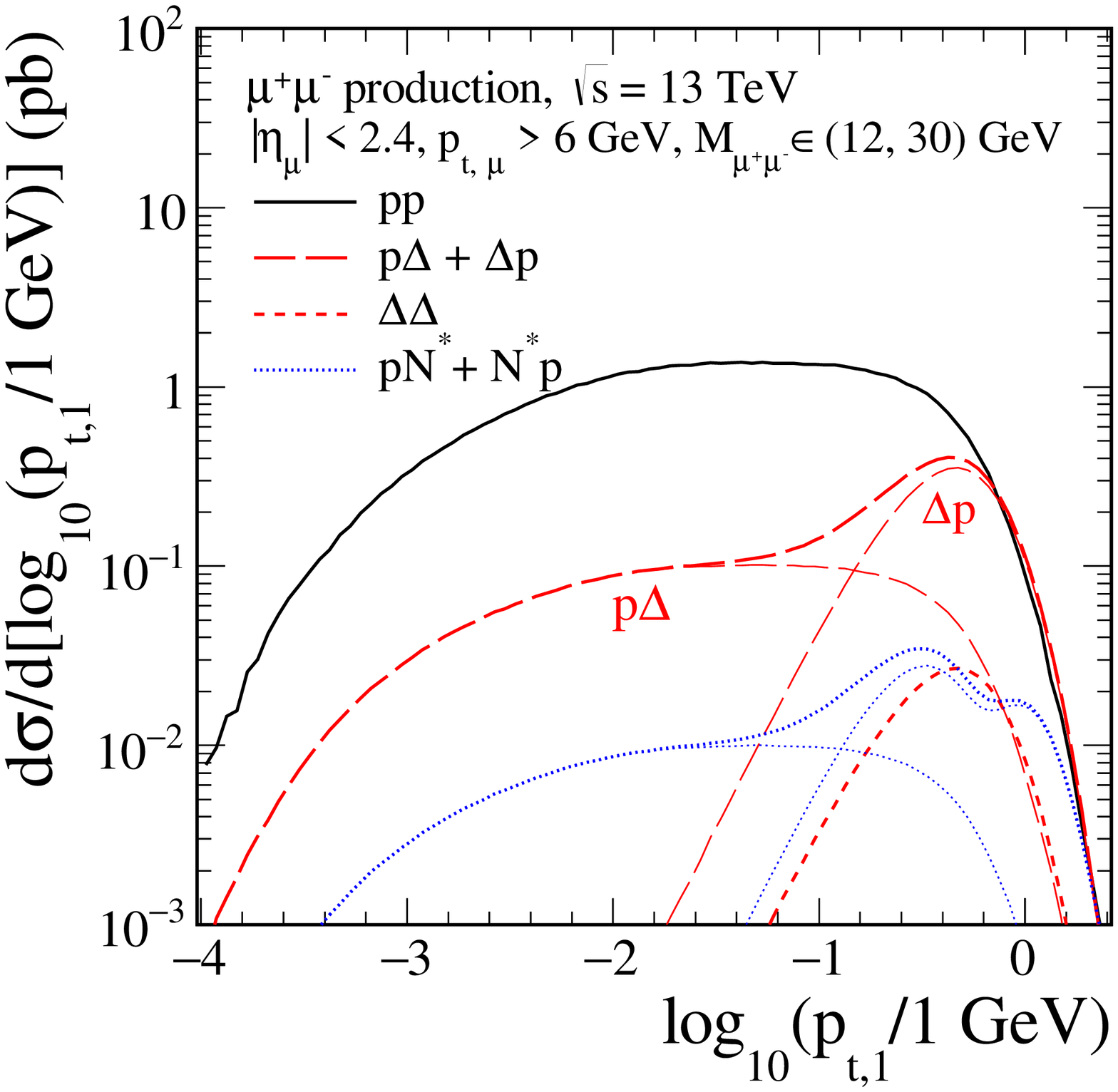}
(b)\includegraphics[width=0.46\textwidth]{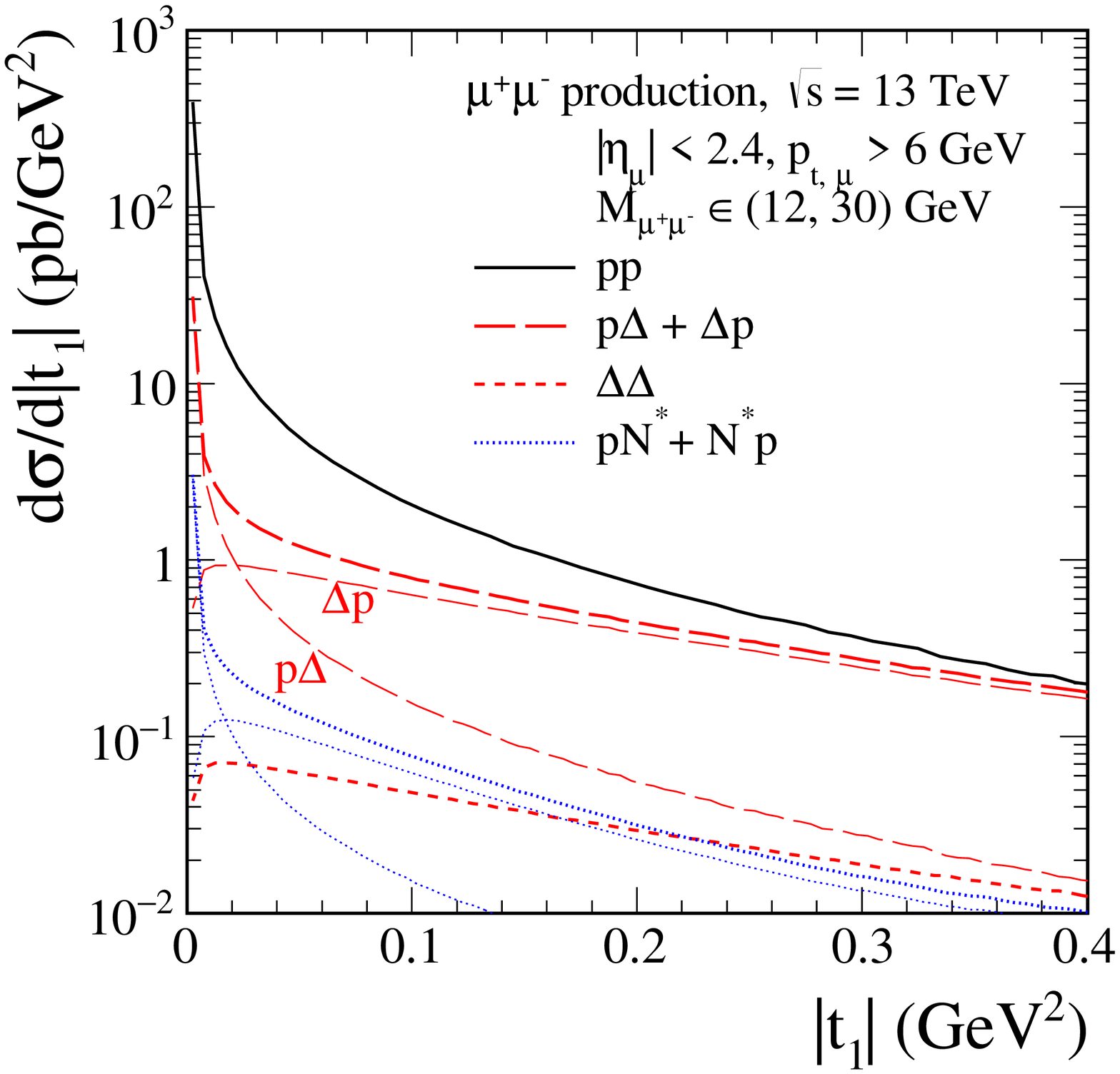}
\caption{\label{fig:1a}
\small
The differential cross sections 
$d\sigma/d[\log_{10}(p_{t,1}/1\,{\rm GeV})]$ (panel (a)), and
$d\sigma/d|t_{1}|$ (panel (b)) for various exclusive processes of the $\mu^{+}\mu^{-}$ production 
at $\sqrt{s} = 13$~TeV and for the ATLAS experimental cuts.
No absorption effects were taken into account here.
}
\end{figure}

In calculating the cross section of the photon initiated dimuon production 
we perform integration in auxiliary variables $\log_{10}(p_{t,1}/1\,{\rm GeV})$
and $\log_{10}(p_{t,2}/1\,{\rm GeV})$ 
instead of the outgoing baryon's transverse momenta ($p_{t,1}$ and
$p_{t,2}$) as usually done. 
The differential distribution $d\sigma/d[\log_{10}(p_{t,1}/1\,{\rm GeV})]$ 
is shown in the panel (a) of 
Fig.~\ref{fig:1a} for the ATLAS kinematics 
($\sqrt{s} = 13$~TeV, $|\eta_{\mu}|<2.5$, $p_{t,\mu} > 6$~GeV, 
$M_{\mu^+ \mu^-} \in (12,30)$~GeV).
Results for the $pp \to pp \mu^+ \mu^-$ reaction (\ref{2to4_reaction_pp})
are shown by the black solid lines.
Results for the reactions (\ref{2to4_reaction_pD}) - (\ref{2to4_reaction_DD})
with $\Delta(1232)$ isobars in the final state are shown by the red dashed lines. 
The blue dotted lines correspond to the contributions with production of 
the Roper resonance $N^{*} \equiv N(1440)$.
In the panel (b) of Fig.~\ref{fig:1a} we present distribution
in transferred four-momentum squared $t_{1}$ between the initial and final baryons.
The distributions in $\log_{10}(p_{t,2}/1\,{\rm GeV})$ and $|t_{2}|$ are the same
as those from the panel (a) and (b), respectively,
but with a different component designations
$\Delta p \leftrightarrow p \Delta$ and $N^{*} p \leftrightarrow p N^{*}$.
We find that from the side of $p \to \Delta(1232)$ and 
$p \to N(1440)$ transitions the differential cross sections $d\sigma/d|t|$ vanish when $|t| \to 0$.

%-----------------------------------------------------------------
%-----------------------------------------------------------------

%-----------------------------------------------------------------
\begin{figure}[!ht]
(a)\includegraphics[width=0.46\textwidth]{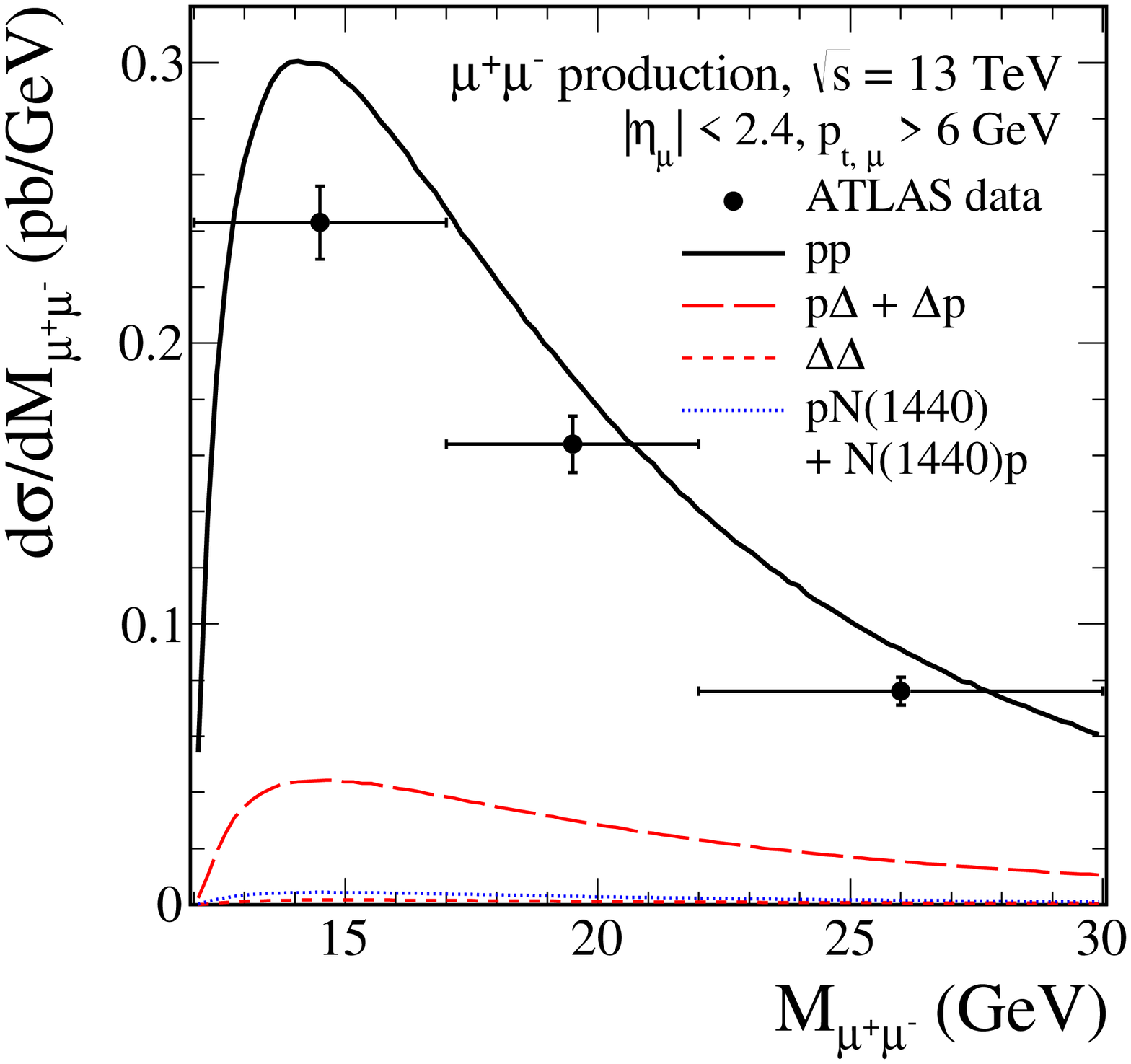}\\
(b)\includegraphics[width=0.46\textwidth]{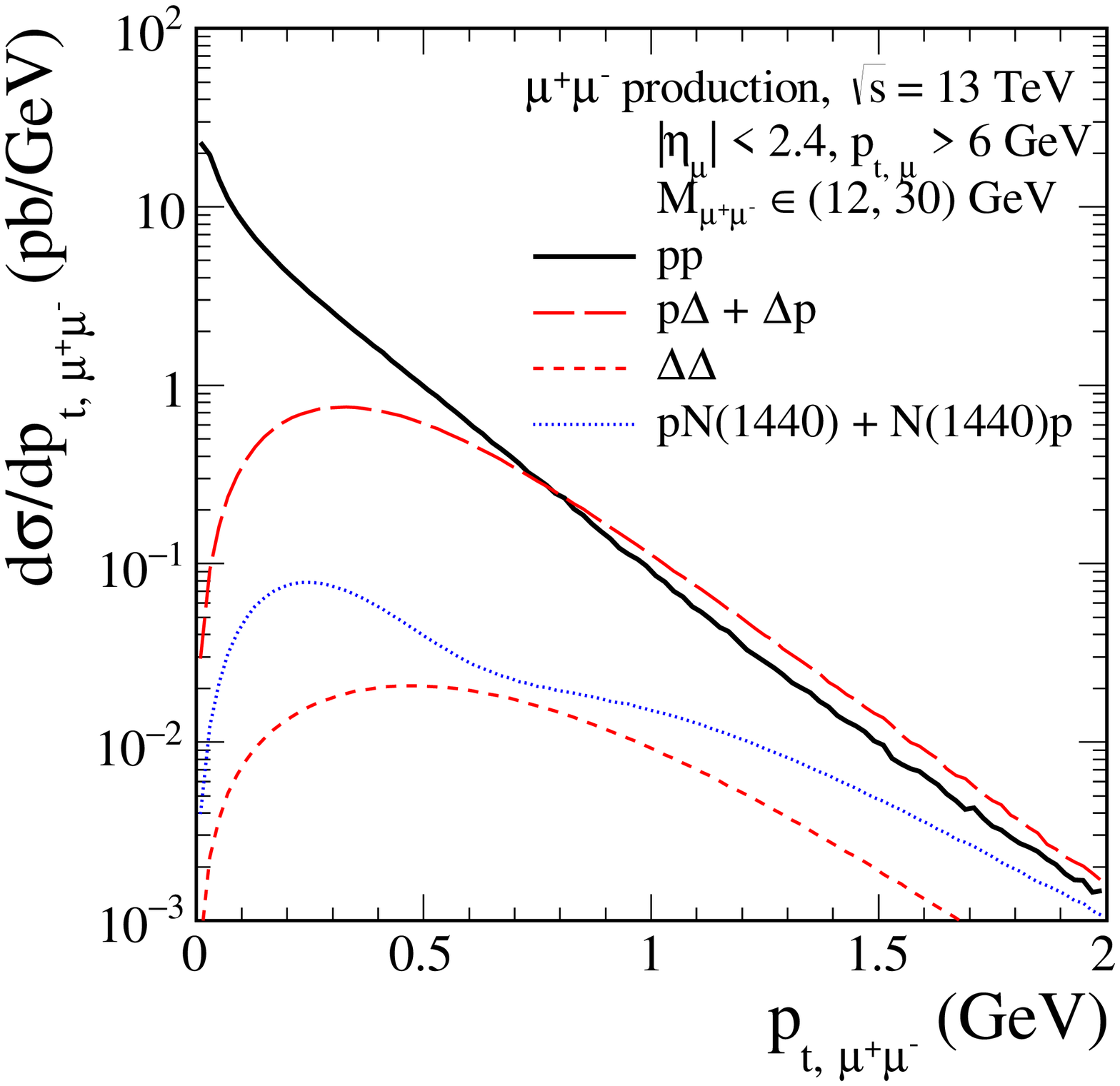}
(c)\includegraphics[width=0.46\textwidth]{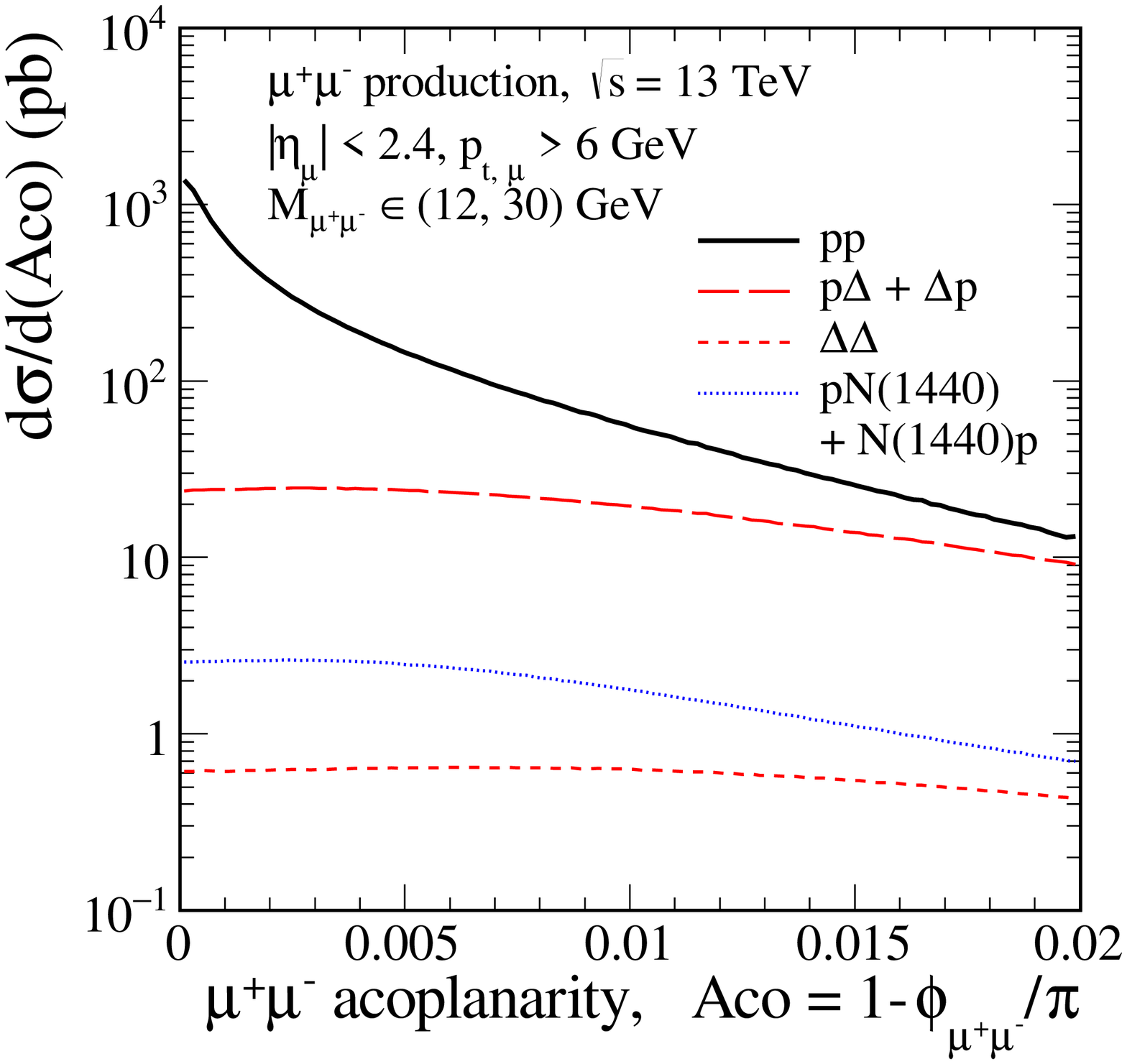}
\caption{\label{fig:1}
\small
The differential cross sections for various exclusive processes
(\ref{2to4_reaction_pp}) - (\ref{2to4_reaction_DD})
specified in the figure legend for the $\mu^{+}\mu^{-}$ production in $pp$ collisions.
Calculations are done for $\sqrt{s} = 13$~TeV, $|\eta_{\mu}|<2.5$, 
$p_{t,\mu} > 6$~GeV, and in dimuon invariant mass region 
$M_{\mu^+ \mu^-} \in (12,30)$~GeV.
No absorption effects are taken into account here.
The ATLAS experimental data from \cite{Aaboud:2017oiq} are shown 
for comparison (see panel (a)).
}
\end{figure}
%-----------------------------------------------------------------

In Fig.~\ref{fig:1} we present differential observables 
for the recent ATLAS experimental cuts~\cite{Aaboud:2017oiq}.
In the panel (a) we show $\mu^+ \mu^-$ invariant mass distributions
for the reactions (\ref{2to4_reaction_pD}) - (\ref{2to4_reaction_DD}).
Here the horizontal error bars mean just bin width.
We see that the $pp \to pp \mu^+ \mu^-$ contribution alone 
(see the black solid line)
exceeds the ATLAS data from \cite{Aaboud:2017oiq}.
As we will show below this is also true 
when including the absorptive corrections in our calculations, see Fig.~\ref{fig:comparison}.
Inclusion of exclusive channels with $\Delta(1232)$ and $N(1440)$ resonances
increases further the cross section for $\mu^+ \mu^-$ production.
In the panel (b) we show distribution in the modulus of sum of the
transverse momentum vectors of muons
$p_{t,\mu^+ \mu^-} = |\bp_{t,\mu^+ \mu^-}|$, $\bp_{t,\mu^+ \mu^-} = \bp_{t,\mu^+} + \bp_{t,\mu^-}$.
For the contributions with $\Delta$ and $N(1440)$ resonances
the cross section $d\sigma/dp_{t,\mu^+ \mu^-}$ vanishes when $p_{t,\mu^+ \mu^-} \to 0$.
The ATLAS experiment imposes a cut on $p_{t,\mu^+ \mu^-} < 1.5$~GeV 
(see Fig.~2 (d) in \cite{Aaboud:2017oiq}).
We can see that such a cut practically does not influence the cross section.
The relative contribution of resonance production
increases with $p_{t,\mu^+ \mu^-}$ and can be even bigger 
than for the $pp \to pp \mu^+ \mu^-$ contribution.
The panel (c) shows the distribution in the dimuon acoplanarity variable
defined by ${\rm Aco} = 1-\phi_{\mu^+ \mu^-}/\pi$, where $\phi_{\mu^+ \mu^-}$
is azimuthal angle between the muons.
Rather different acoplanarity distributions are obtained
for the different processes considered here.
However, no acoplanarity cut is imposed by the recent ATLAS experiment.

%-----------------------------------------------------------------
\begin{figure}[!ht]
\includegraphics[width=0.49\textwidth]{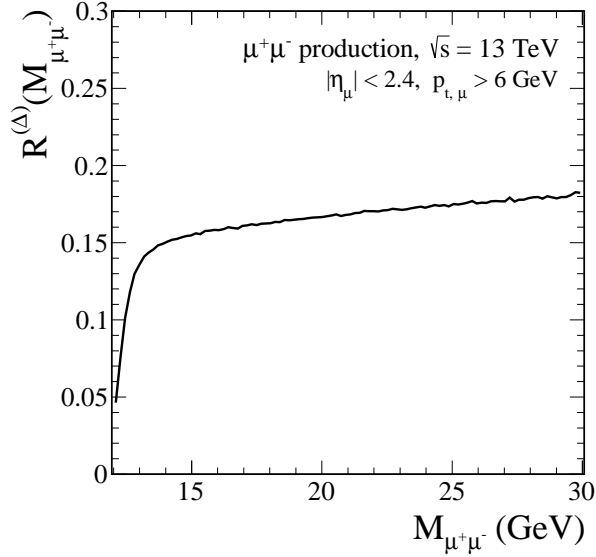}
\caption{\label{fig:2}
\small
The ratio $R^{(\Delta)}(M_{\mu^+ \mu^-})$ defined by
Eq.(\ref{ratio_Delta}) for the $\mu^{+}\mu^{-}$ production
at $\sqrt{s} = 13$~TeV and for the ATLAS cuts.
}
\end{figure}
%-----------------------------------------------------------------

In Fig.~\ref{fig:2} we show the ratio
\begin{equation}
R^{(\Delta)}(M_{\mu^+ \mu^-}) = \frac{d\sigma^{(p \Delta)} / dM_{\mu^+ \mu^-} + 
                                    d\sigma^{(\Delta p)} / dM_{\mu^+ \mu^-} + 
                                    d\sigma^{(\Delta \Delta)} / dM_{\mu^+ \mu^-}}
                                   {d\sigma^{(pp)} / dM_{\mu^+ \mu^-}}
\label{ratio_Delta}
\end{equation}
for $\sqrt{s} = 13$~TeV and the ATLAS experimental cuts.
In (\ref{ratio_Delta}), e.g., $d\sigma^{(p \Delta)} / dM_{\mu^+ \mu^-}$ 
is the differential cross section
for the $pp \to p  \mu^{+} \mu^{-} \Delta^{+}$ reaction (\ref{2to4_reaction_pD}).
The contribution of the new processes (\ref{2to4_reaction_pD}) - (\ref{2to4_reaction_DD})
increases with increasing $M_{\mu^+ \mu^-}$.
The ratio exceeds 15\% for $M_{\mu^+ \mu^-} > 14$~GeV.

So far we have omitted effect related to extra soft interactions
which lead to a reduction of the cross section with the extra requirement of rapidity gap.
How big is effect of the absorption associated with different exclusive effects?
Is it the same for different components? These questions are very
important but go beyond the scope of the present paper.

\begin{figure}[!ht]
(a)\includegraphics[width=0.46\textwidth]{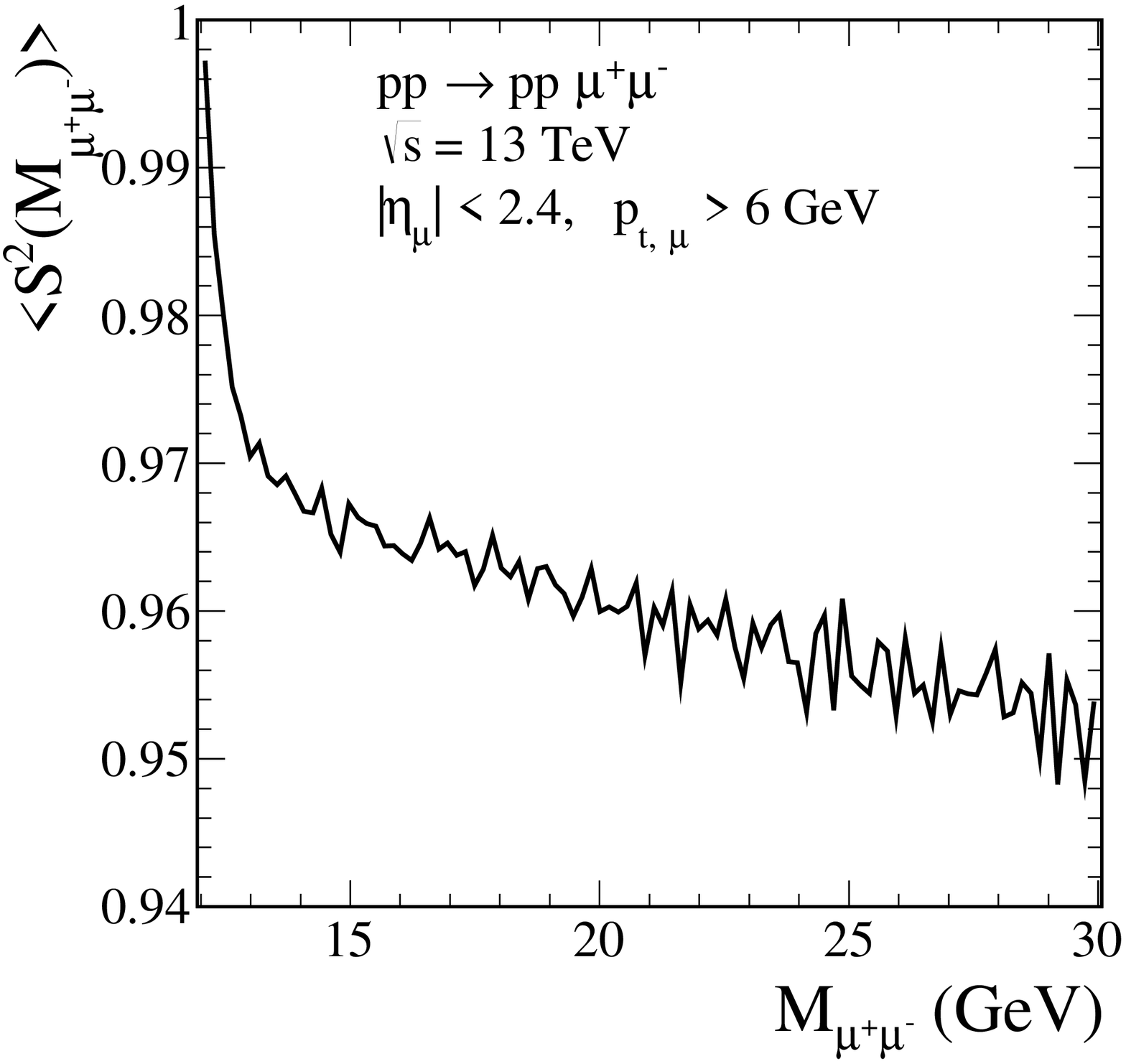}\\
(b)\includegraphics[width=0.46\textwidth]{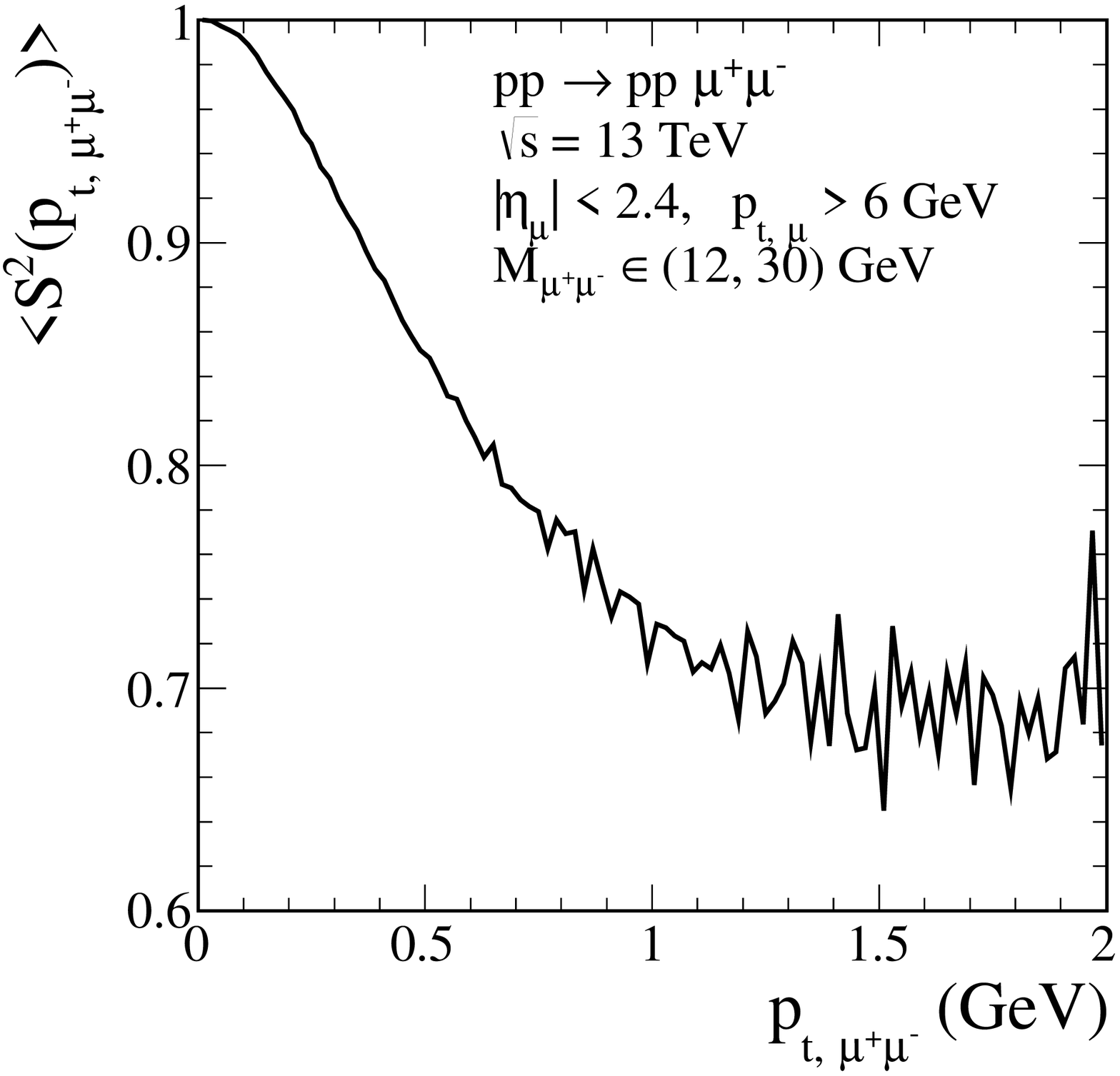}
(c)\includegraphics[width=0.46\textwidth]{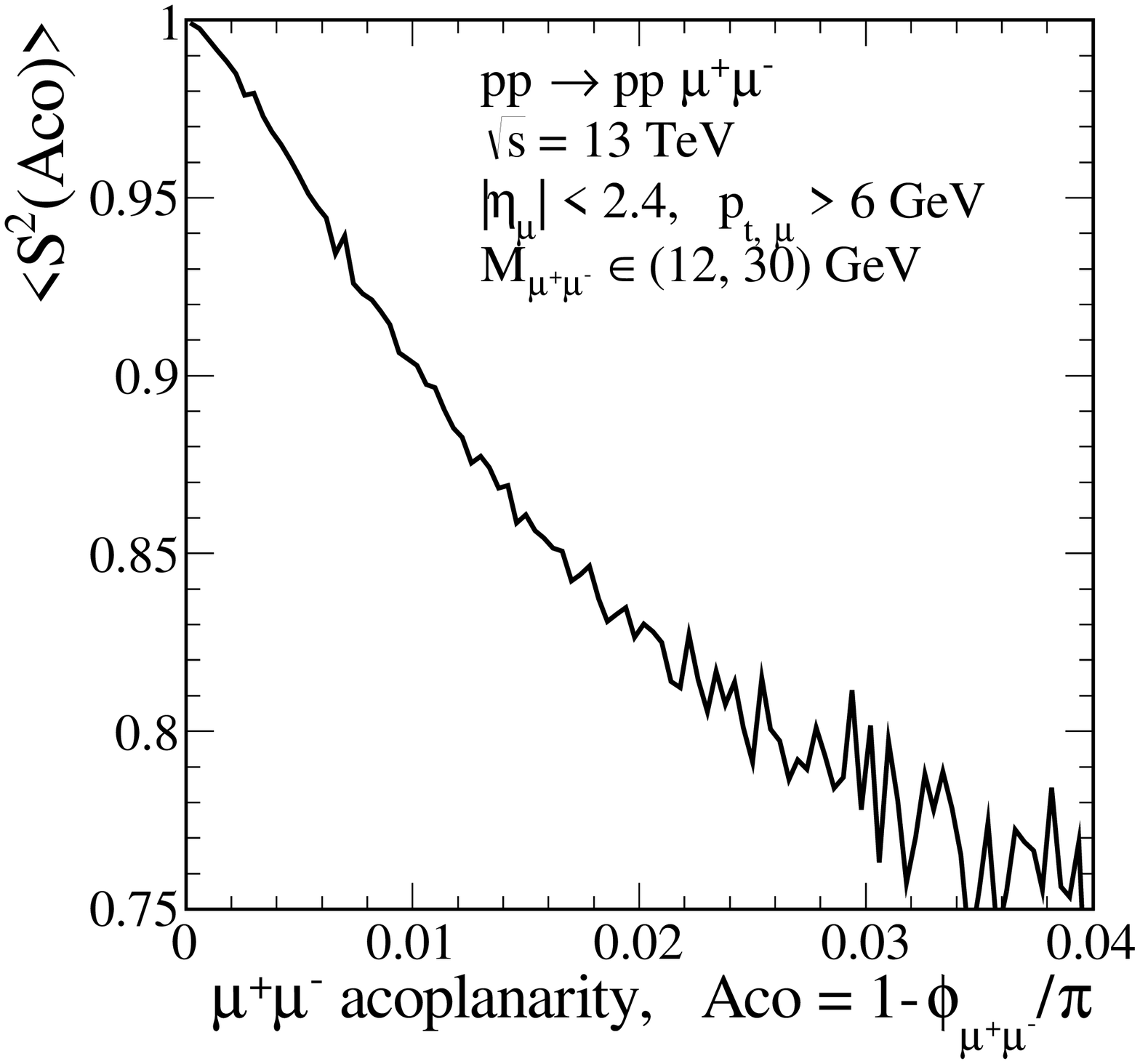}
\caption{\label{fig:3}
\small
The ratios $\langle S^{2} \rangle$ (\ref{R_abs}) as a function of
$M_{\mu^+ \mu^-}$ (panel (a)), 
$p_{t,\mu^+ \mu^-}$ (panel (b)), dimuon acoplanarity (panel (c)).
%, and $|t_{1}|$ (the panel (d)).
Calculations were done at $\sqrt{s} = 13$~TeV and for the ATLAS cuts.
}
\end{figure}

In Fig.~\ref{fig:3} we show the relative effect of absorption,
for the $pp \to pp \mu^+ \mu^-$ reaction,
\begin{equation}
\langle S^{2}(x) \rangle = \frac{d\sigma^{{\rm absorption}}/ dx}{d\sigma^{{\rm Born}} /dx}\,,
\label{R_abs}
\end{equation}
where $x = M_{\mu^+ \mu^-}$, $p_{t,\mu^+ \mu^-}$, ${\rm Aco}$.
In (\ref{R_abs}) $d\sigma^{{\rm absorption}}/dx$ is the differential cross section 
including the absorptive effects at the amplitude level as described in Sec.~\ref{absorption}
and $d\sigma^{{\rm Born}}/dx$ is the differential cross section without the absorption.
We predict somewhat larger absorption for 
the $p N^*$, $N^* p$, and $N^* N^*$ contributions than
for the traditional $pp$ final state (not shown here).
Our calculations suggest similar effect for the $p \Delta$, $\Delta p$,
and $\Delta \Delta$ final states for which explicit calculation
is rather difficult.
The effect of the absorption depends on kinematics but is rather small. 
The effect would increase somewhat when adding intermediate proton 
resonance states. 
Such a calculation is more difficult and requires input,
$p N^* \Pom$, $p N^* \gamma$ couplings, which is 
not available at present.

%-----------------------------------------------------------------
%-----------------------------------------------------------------

%-----------------------------------------------------
\subsection{$k_t$-factorization approach}
\label{sec:kt_approach}
%-----------------------------------------------------

In this subsection we wish to show the differential distributions
obtained in the $k_t$-factorization approach.
The results of the single or double dissociative processes enter
the cross section via so-called deep-inelastic structure functions.
%Only $F_2$ structure function will be included here.
In our calculation we have used two different parametrizations 
of the proton structure function $F_{2}(x,Q^{2})$ taken from the literature:
(1) Fiore \textit{et al.} parametrization \cite{Fiore:2002re,Fiore:2003dg}
(labeled by us FFJLM) based on a Regge-dual model
that explicitly includes the prominent nucleon resonances plus a smooth background
to describe the $F_2$ experimental data for $p(e,e')X$ reaction measured at the JLab;
and (2) Szczurek-Uleshchenko (SU) parametrization \cite{Szczurek:1999rd}
which gives good description at rather small and intermediate $Q^{2}$ at not too small $x$.

\begin{figure}[!ht]
(a)\includegraphics[width=0.46\textwidth]{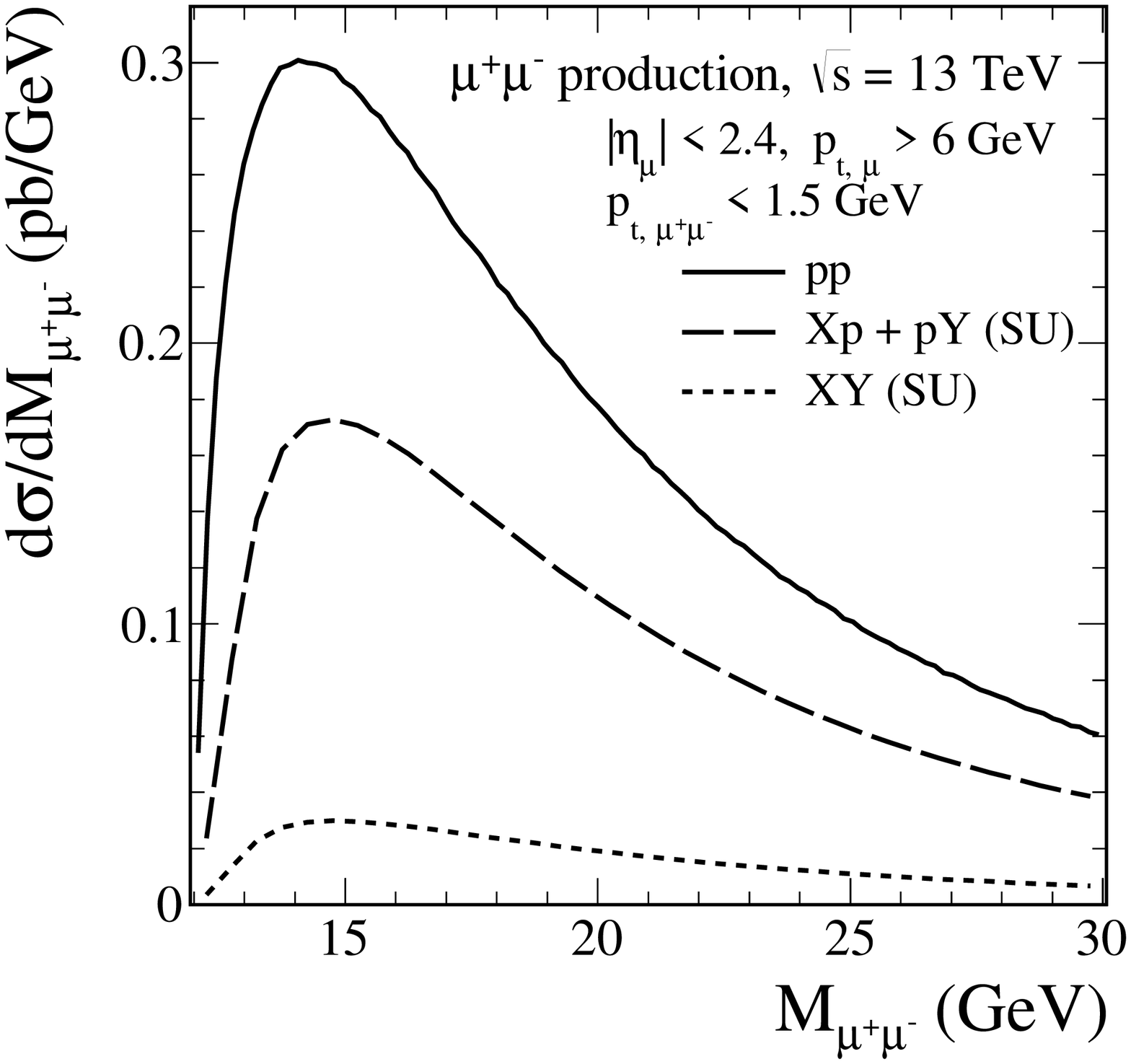}
(b)\includegraphics[width=0.46\textwidth]{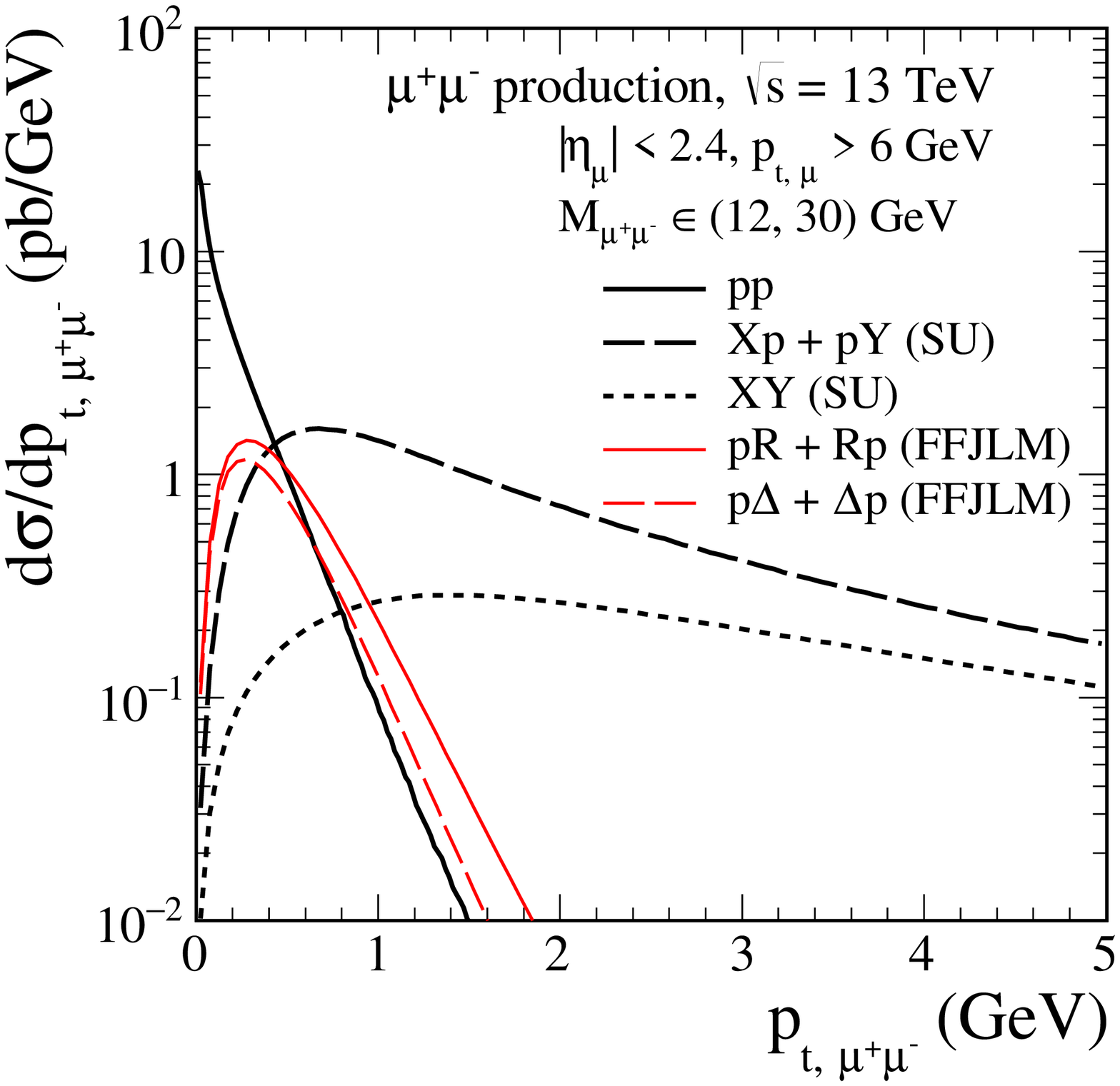}\\
(c)\includegraphics[width=0.46\textwidth]{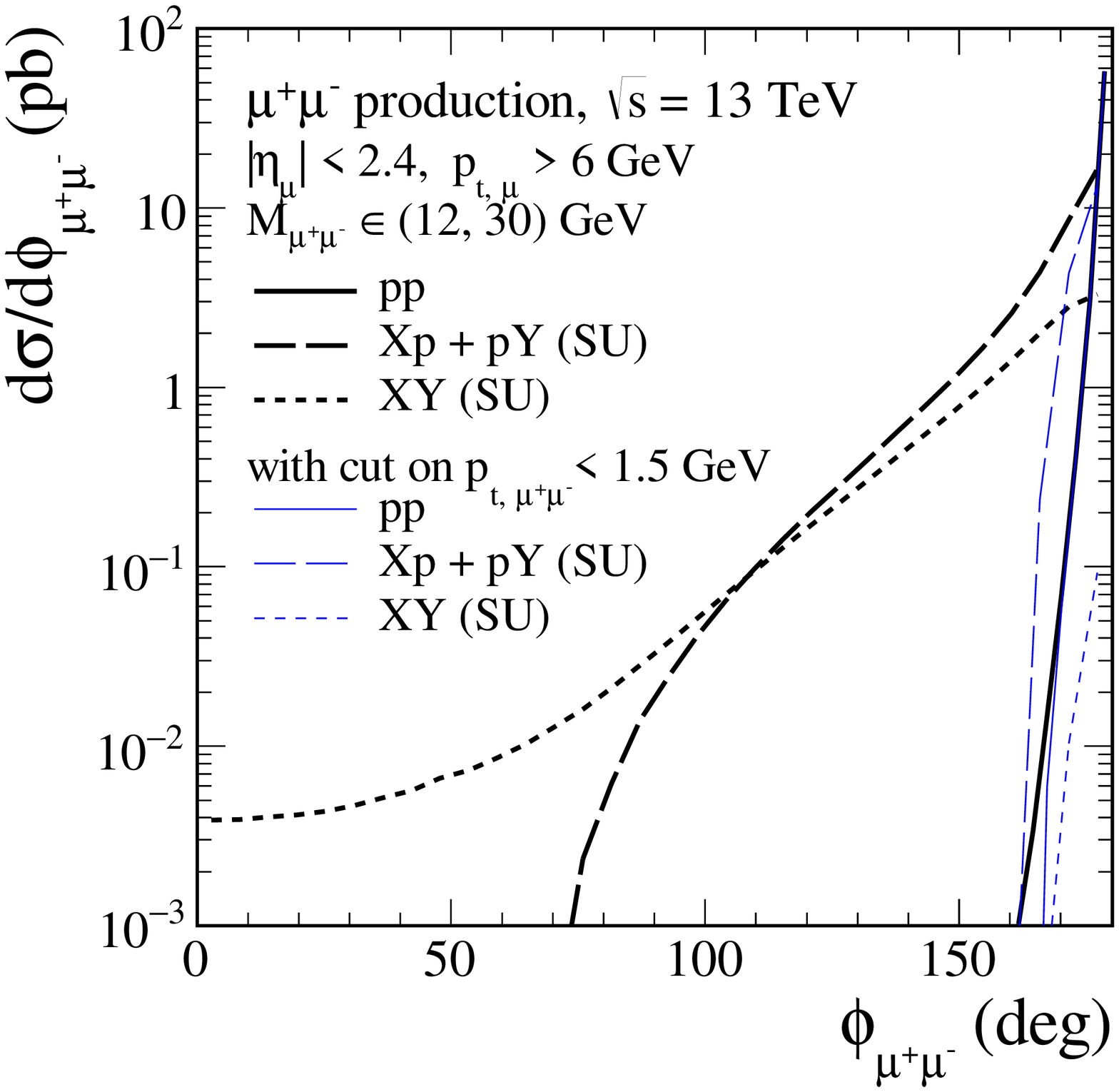}
  \caption{\label{fig:kt_fact_SU}
  \small
Single and double dissociative continuum contributions for 
$d\sigma/dM_{\mu^+ \mu^-}$ (panel (a)),
$d\sigma/dp_{t,\mu^+ \mu^-}$ (panel (b)), 
and $d\sigma/d\phi_{\mu^+ \mu^-}$ (panel (c))
obtained within the $k_t$-factorization approach
for the recent ATLAS experimental cuts \cite{Aaboud:2017oiq}.
For the continuum processes the Szczurek-Uleshchenko parametrization (labeled as SU) 
was used in the calculation and we impose an upper limit on dissociative systems $M_X, M_Y < 50$~GeV.
For comparison, the solid lines represent the results for the purely elastic contribution (\ref{2to4_reaction_pp}).
In the panel (b) also the results for resonant contributions (labeled as FFJLM) are shown.
In the panel (c) the black lines correspond to the results 
without the cut on $p_{t,\mu^+ \mu^-}$.
The blue lines show results obtained including all experimental cuts.}
\end{figure}

In Fig.~\ref{fig:kt_fact_SU} we compare different distributions
($d\sigma/dM_{\mu^+ \mu^-}$, $d\sigma/dp_{t,\mu^+ \mu^-}$, $d\sigma/d\phi_{\mu^+ \mu^-}$) 
for purely elastic (the solid line), single dissociative (the dashed line), 
and double dissociative (the dotted line) contributions. 
In the calculations we have included all cuts of the ATLAS experiment \cite{Aaboud:2017oiq}, 
including also the cut on $p_{t,\mu^+ \mu^-} < 1.5$~GeV.
Results for the continuum dissociative contributions (labeled as SU)
were calculated using the Szczurek-Uleshchenko parametrization of $F_2$ experimental data.
The shapes of the $M_{\mu^+ \mu^-}$ distributions (see the panel (a)) 
are very similar while the other distributions are rather different.
In the panel (b) we present in addition results for resonance production
obtained in the FFJLM parametrization; see the red lines (labeled as FFJLM).
Here $pR + Rp$ denotes contribution when
three resonances $\Delta(1232)$ $\frac{3}{2}^{+}$, 
$N^{*}(1520)$ $\frac{3}{2}^{-}$, and $N^{*}(1680)$ $\frac{5}{2}^{+}$ 
are added together.
For comparison, we show also the result with only $\Delta(1232)$ resonances 
($p \Delta + \Delta p$ component).
One can explicitly see that the cut on $p_{t,\mu^+ \mu^-}$ reduces
the dissociative contributions (see also Table~\ref{tab:table1}).
No cut on $\phi_{\mu^+ \mu^-}$ was imposed in the ATLAS experiment.
But it should be realized that such a cut is strongly correlated with the cut on $p_{t,\mu^+ \mu^-}$
(compare the black and blue lines in the panel (c)).
%--------------------------------------------------------
%--------------------------------------------------------

We remind that the ATLAS collaboration imposes extra condition 
on $p_{t,\mu^+ \mu^-} < 1.5$~GeV \cite{Aaboud:2017oiq}.
Inclusion of such a cut suppresses the relative amount of dissociative
continuum contributions but definitely does not solve the problem of the need
of large absorption effects.
A large size of the dissociation into continuum requires a special comment.
At large $p_{t,\mu^+ \mu^-}$ the single and double continuum
dissociative processes should dominate.
As shown recently in \cite{Forthomme:2018sxa} the absorption effect for
$\sqrt{s} = 13$~TeV and $M_X, M_Y < 50$~GeV associated with remnant
fragmentation(s) are rather small.
So far other absorption effects were not calculated consistently in the literature.
% and cannot explain why their contribution is ``not seen'' 
%by the ATLAS experiment \cite{Aaboud:2017oiq}.
%We expect that these contributions are
%subjected to much larger absorption effect. 
%This is by itself interesting topic which goes, however, 
%beyond of the scope of the present paper where
%we focus on production of proton resonances.

At present ATLAS collaboration used some procedure to reduce the 
background from the single and double dissociation processes
(see Sec.~5 of \cite{Aaboud:2017oiq}).
However, this procedure may be model dependent 
and in our opinion requires further studies. 
It would be valuable to confront the extracted contribution 
with our model calculation.
Also absorption effects for the continuum dissociation
are not fully understood in our opinion.

%-------------------------------------------------------------------------
\subsection{Cross sections and comparison with the ATLAS experimental data} 
\label{sec:Cross_sections}
%-------------------------------------------------------------------------

The ATLAS collaboration has measured the fiducial cross section
of the $pp \to pp \mu^{+} \mu^{-}$ reaction at $\sqrt{s} = 13$~TeV \cite{Aaboud:2017oiq}.
The experimental result is
\begin{eqnarray}
\sigma_{{\rm exp., \,fid.}}(pp \to pp \mu^{+} \mu^{-}) = 3.12 \pm 0.07\;({\rm stat.}) \pm 0.14\;({\rm syst.})\; {\rm pb}\,,
\label{cs_ATLAS}
\end{eqnarray}
for both dimuon invariant mass ranges and for $p_{t, \mu}$ and 
$|\eta_{\mu}|$ requirements:
\begin{eqnarray}
&& 12\;{\rm GeV} < M_{\mu^{+} \mu^{-}} < 30\;{\rm GeV}\,,\; p_{t, \mu} > 6\;{\rm GeV}\,,\; |\eta_{\mu}|<2.4\,,
\label{ATLAS_cuts_A}\\
&& 30\;{\rm GeV} < M_{\mu^{+} \mu^{-}} < 70\;{\rm GeV}\,,\; p_{t, \mu} > 10\;{\rm GeV}\,\;, |\eta_{\mu}|<2.4\,.
\label{ATLAS_cuts_B}
\end{eqnarray}

The sum of cross sections calculated within the ``exact $2 \to 4$ approach'' (see Sec.~\ref{exact})
for the experimental cuts (\ref{ATLAS_cuts_A}) and (\ref{ATLAS_cuts_B}), respectively, is found to be
\begin{eqnarray}
\sigma_{{\rm exact}}^{({\rm Born})}(pp \to pp \mu^{+} \mu^{-}) = 
3.01\;{\rm pb} + 0.55\;{\rm pb} = 3.56\;{\rm pb}\,
\label{cs_exact_Born}
\end{eqnarray}
without the absorptive corrections, and 
\begin{eqnarray}
\sigma_{{\rm exact}}^{({\rm absorption})}(pp \to pp \mu^{+} \mu^{-}) = 
2.89\;{\rm pb} + 0.51\;{\rm pb} = 3.40\;{\rm pb}\,
\label{cs_exact_abs}
\end{eqnarray}
including the absorptive corrections as discussed in Sec.~\ref{absorption}.

The authors of \cite{Aaboud:2017oiq} compare their result (\ref{cs_ATLAS})
with the theoretical predictions of two models with absorptive corrections.
Our result (\ref{cs_exact_abs}) is in good agreement with 
the SuperChic MC \cite{Harland-Lang:2015cta} result
$\sigma = 3.45 \pm 0.05$~pb quoted in \cite{Aaboud:2017oiq}.
However, smaller cross section was obtained in the finite-size EPA approach \cite{Dyndal:2014yea} 
that gives $\sigma = 3.06 \pm 0.05$~pb.

%-----------------------------------------------------------------
\begin{figure}[!ht]
\includegraphics[width=0.5\textwidth]{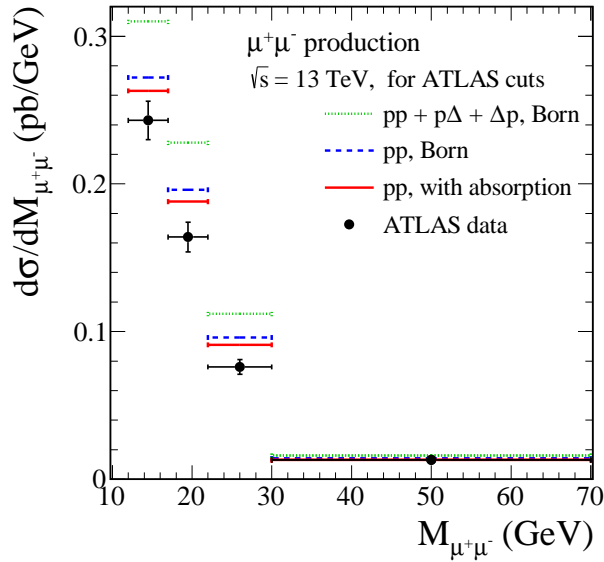}
\caption{\label{fig:comparison}
\small
The differential cross sections $d\sigma/dM_{\mu^{+} \mu^{-}}$
for the $\mu^{+}\mu^{-}$ production at $\sqrt{s} = 13$~TeV with the
ATLAS experimental cuts 
specified in (\ref{ATLAS_cuts_A}) and (\ref{ATLAS_cuts_B}).
Our ``exact $2 \to 4$ kinematics'' predictions (lines) are compared 
with the ATLAS differential fiducial cross sections from Table~3 
of \cite{Aaboud:2017oiq}.
The green-dotted lines and blue-dashed lines represent the Born results for
the $pp \to pp \mu^{+} \mu^{-}$ plus 
that for the $pp \to p\Delta \mu^{+} \mu^{-} (\Delta p \mu^{+} \mu^{-})$ processes
and for the $pp \to pp \mu^{+} \mu^{-}$ process alone, respectively.
The red solid bottom lines represent the results for the 
$pp \to pp \mu^{+} \mu^{-}$ reaction with the absorptive effects included.
}
\end{figure}
%-----------------------------------------------------------------

In Fig.~\ref{fig:comparison} we present the dimuon invariant mass distributions
for our ``exact $2 \to 4$ kinematics'' approach for the $pp \to p p \mu^+ \mu^-$ process, 
without (the blue dashed lines) and with (the red solid lines) absorption effects
together with the ATLAS results from Table~3 of \cite{Aaboud:2017oiq}.
For the $pp \to pp \mu^{+} \mu^{-}$ component
we get the theoretical survival factor values 
$\langle S^{2} \rangle = 
\sigma_{{\rm exact}}^{({\rm absorption})}/\sigma_{{\rm exact}}^{({\rm Born})} = 0.97$, 0.96, 0.95, and 0.93
integrated according to the four experimental bins, respectively.
For comparison we need 0.92, 0.87, 0.84, and 1.00 to describe the ATLAS data.
Larger suppression factors are necessary to describe the ATLAS data
especially when taking into account contributions with $\Delta$ isobars 
in the final state: 0.78, 0.72, 0.68, and 0.81, respectively.
These numbers are far from unity often naively expected for two-photon exclusive processes.
In general, the absorption effects may depend on the final state.
Here we have presented our estimates for the $p p \to p p \mu^+ \mu^-$ process. 
As already discussed, it is very difficult to make similar
predictions for the $p p \to p \Delta \mu^+ \mu^-$ or
$p p \to \Delta p \mu^+ \mu^-$ processes.
A comparison with experimental data suggests that the corresponding
effects should be much bigger than for the $p p \to p p \mu^+ \mu^-$ process.

%----------------------------------------------------------------------------
\begin{table}
\begin{small}
\caption{
Cross sections for different processes
for central exclusive production of $\mu^{+} \mu^{-}$ pairs 
calculated for three different approaches.
The calculations was performed for $\sqrt{s} = 13$~TeV and with different experimental cuts.
In the $k_{t}$-factorization approach for the continuum processes (labeled as SU) 
we take an upper limit on $M_X$ and $M_Y < 50$~GeV.
In the case of resonance production (FFJLM) $R$ means processes when 
contributions of three resonances $\Delta(1232)$, $N^{*}(1520)$, and
$N^{*}(1680)$ are added together.
No absorption effects were included here.
}
\label{tab:table1}
\begin{center}
\begin{tabular}{|l|r|r|r|r|r|}
\hline
$|\eta_{\mu}| < 2.4$                & Y            & Y             & Y             & Y \\
$p_{t,\mu}> 6$~GeV                  &              & Y             & Y             & Y \\
$12<M_{\mu^{+}\mu^{-}}<30$~GeV      &              &               & Y             & Y \\
$p_{t,\mu^{+}\mu^{-}} < 1.5$~GeV    &              &               &               & Y \\
\hline \hline 
\textbf{Exact $2 \to 4$ approach}&\textbf{$\sigma$ (nb)} & \textbf{$\sigma$ (pb)} & \textbf{$\sigma$ (pb)} & \textbf{$\sigma$ (pb)} \\
\hline  
$p p \to p p \,\mu^{+} \mu^{-}$               &32.56 & 3.81 & 3.01 & 3.01\\
$p p \to p \Delta \,\mu^{+} \mu^{-}$          & 0.67 & 0.31 & 0.23 & 0.23\\  
$p p \to \Delta p \,\mu^{+} \mu^{-}$          & 0.67 & 0.31 & 0.23 & 0.23\\          
$p p \to \Delta \Delta \,\mu^{+} \mu^{-}$     & 0.02 & 0.02 & 0.02 & 0.02\\
%$p p \to p N(1440) \,\mu^{+} \mu^{-}$          &  &    &   & \\ 
%$p p \to N(1440) p \,\mu^{+} \mu^{-}$          &  &    &   & \\
%$p p \to N(1440) N(1440) \,\mu^{+} \mu^{-}$          &  &    &    & \\  
\hline \hline
\textbf{EPA} &              &               &               &  \\
\hline
$p p \to p p \,\mu^{+} \mu^{-}$            &37.08 & 3.68 & 2.97 & \\
$p p \to p \Delta \,\mu^{+} \mu^{-}$       & 1.87 & 0.33 & 0.26 & \\
$p p \to \Delta p \,\mu^{+} \mu^{-}$       & 1.87 & 0.33 & 0.26 & \\
$p p \to \Delta \Delta \,\mu^{+} \mu^{-}$  & 0.09 & 0.03 & 0.02 & \\
\hline \hline
\textbf{$k_{t}$-factorization approach}     &              &               &               &  \\
\hline      
$p p \to p p \,\mu^{+} \mu^{-}$                      &39.74 & 3.91 & 3.16 & \\
$p p \to p \Delta \,\mu^{+} \mu^{-}$      (FFJLM)    & 1.33 & 0.41 & 0.32 & \\
$p p \to \Delta p \,\mu^{+} \mu^{-}$      (FFJLM)    & 1.33 & 0.41 & 0.32 & \\ 
$p p \to \Delta \Delta \,\mu^{+} \mu^{-}$ (FFJLM)    & 0.02 & 0.01 &  & \\  
\hline         
$p p \to p R \,\mu^{+} \mu^{-}$ (FFJLM)              & 1.65 &  0.55 & 0.43 & \\
$p p \to R p \,\mu^{+} \mu^{-}$ (FFJLM)              & 1.65 &  0.55 & 0.43 & \\
$p p \to R R \,\mu^{+} \mu^{-}$ (FFJLM)              & 0.03 &  0.02 &  & \\  
\hline        
$p p \to p Y \,\mu^{+} \mu^{-}$ (SU)    &  &  2.38  &   1.84 & 0.88 \\
$p p \to X p \,\mu^{+} \mu^{-}$ (SU)    &  &  2.38  &   1.84 & 0.88\\
$p p \to X Y \,\mu^{+} \mu^{-}$ (SU)    &  &  1.76  &   1.32 & 0.30\\
\hline
\end{tabular}
\end{center}
\end{small}
\end{table}
%----------------------------------------------------------------------------

In Table~\ref{tab:table1} we have collected integrated cross sections 
for different contributions calculated in three different approaches:
exact $2 \to 4$ (see Secs.~\ref{exact} and \ref{absorption}), EPA (see Sec.~\ref{EPA}), 
and $k_{t}$-factorization (see Sec.~\ref{kt_approach}).
Results for experimental cuts $|\eta_{\mu}| < 2.4$, 
$p_{t,\mu}$, $M_{\mu^{+}\mu^{-}}$, and $p_{t, \mu^{+}\mu^{-}}$ are shown.
The results obtained within the exact $2 \to 4$
approach imposing the ATLAS cuts (\ref{ATLAS_cuts_A}) 
are similar to the results obtained within the EPA approach.
We get a slightly larger cross section for the $p p \to p p \mu^+ \mu^-$ process
within the $k_{t}$-factorization approach.
There are also results for the single and double resonance production 
(FFJLM parametrization \cite{Fiore:2002re})
and dissociative continuum contributions 
(SU parametrization \cite{Szczurek:1999rd})
calculated within the $k_{t}$-factorization approach.
The resonance production constitutes 
about 20\% of the fully exclusive $p p \to p p \mu^+ \mu^-$ contribution,
that is, somewhat larger than from our $2 \to 4$ calculation without absorption effects.
The double resonance contribution is less than 1\% of the
purely exclusive component and can be in practice neglected.
The contributions of single and double dissociation continuum 
with all the cuts described in Table~\ref{tab:table1}
constitute 68\% of the purely exclusive component
but was hopefully removed by the ATLAS extraction procedure \cite{Aaboud:2017oiq}. 
These are non-negligible contributions, larger than typical size of
absorption effects for the $p p \to p p \mu^+ \mu^-$ process.

We have much larger problem of overestimating the ATLAS experimental
data than signaled in \cite{Aaboud:2017oiq}, see also Fig.~\ref{fig:comparison}.
This probably means that the continuum contributions are subjected
to much larger absorption effects 
than contributions with resonances in the final state.
% which destroy rapidity gaps. 
This is very interesting problem but clearly goes beyond the scope of 
the present paper, where we have focused mainly on the contributions with
resonances in the final state.
Solving the problem requires probably inclusion of multi-parton processes \cite{Babiarz:2017jxc} 
and remnant fragmentation \cite{Khoze:2017sdd}. 
The parameters of the multi-parton interactions were adjusted rather to 
$gg$ induced processes and cannot be used for our $\gamma \gamma$ 
induced processes.
Recently, in Ref.~\cite{Forthomme:2018sxa}, the effect of rapidity 
gap survival factor associated with remnant fragmentation was studied 
for $W^+ W^-$ production.
Such effects strongly depend on details of experiment.
The effects of absorption were of course not included 
by the ATLAS collaboration when ``subtracting'' the dissociative contributions.

%It may mean that the absorption effects are too small 
%or alternatively that a contamination of the continuum contributions
%is not correctly removed. 
%Both effects should be therefore studied in detail in a future.

%-------------------------------------------------------------------------
\subsection{Predictions for ATLAS + ALFA experiment} 
\label{sec:Predictions_ALFA}
%-------------------------------------------------------------------------

The measurement of forward protons would be useful 
in our opinion to better understand absorption effects.
%In current experiments at the LHC it is of great advantage for the theoretical analysis
%if the leading outgoing protons can be measured.
There are several efforts to complete installation of forward proton detectors.  
The CMS collaboration combines efforts with the TOTEM collaboration 
while the ATLAS collaboration may use the ALFA sub-detectors.

Here we wish to show our predictions for $\sqrt{s} = 13$~TeV based on 
the exact $2 \to 4$ approach (see Sec.~\ref{exact})
including the ATLAS experimental cuts (\ref{ATLAS_cuts_A})
and with extra cuts on the leading protons of 0.17~GeV~$< |p_{y,1}|, |p_{y,2}|<$~0.5~GeV \cite{Adamczyk_priv}
as will be the proton momentum window for ALFA detectors on both sides of the ATLAS detector. 

We obtain the Born cross section
\begin{eqnarray}
\sigma_{{\rm exact}}^{({\rm Born})}(pp \to pp \mu^{+} \mu^{-}) =
12.87\;{\rm fb} + 12.12\;{\rm fb} = 24.99\;{\rm fb}\,
\label{cs_exact_Born_ALFA}
\end{eqnarray}
and including the absorptive corrections (see Sec.~\ref{absorption})
\begin{eqnarray}
\sigma_{{\rm exact}}^{({\rm absorption})}(pp \to pp \mu^{+} \mu^{-}) = 
11.32\;{\rm fb} + 2.56\;{\rm fb} = 13.88\;{\rm fb}\,.
\label{cs_exact_abs_ALFA}
\end{eqnarray}
In (\ref{cs_exact_Born_ALFA}) and (\ref{cs_exact_abs_ALFA}) 
we sum resulting cross sections for two exclusive conditions
$p_{y,1} p_{y,2} < 0$ and $p_{y,1} p_{y,2} \geqslant 0$.
We obtain much bigger reduction of the cross section 
due to absorption effects when $p_{y,1} p_{y,2} \geqslant 0$.

%-----------------------------------------------------------------
\begin{figure}[!ht]
(a)\includegraphics[width=0.46\textwidth]{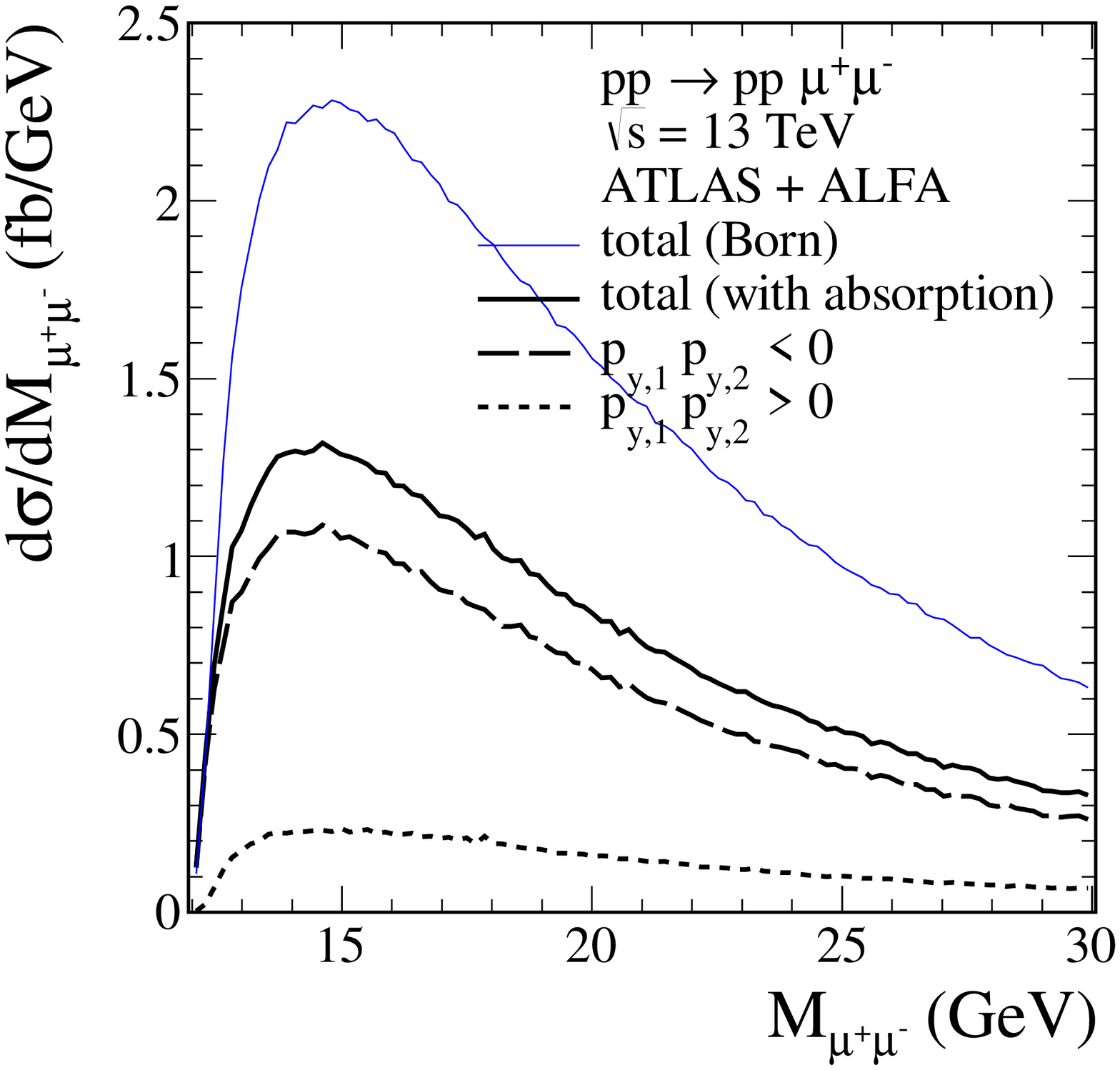}\\
(b)\includegraphics[width=0.46\textwidth]{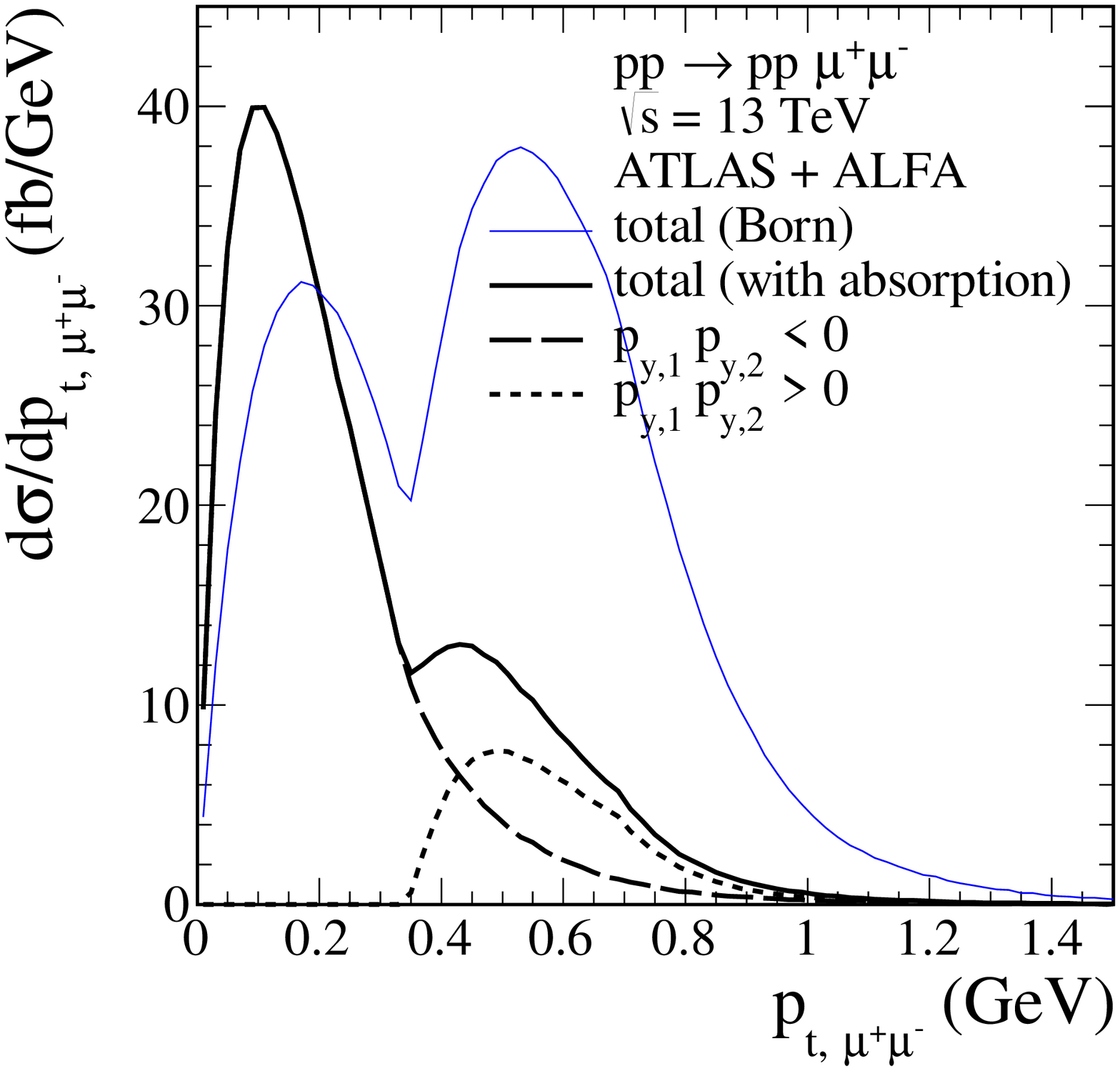}
(c)\includegraphics[width=0.46\textwidth]{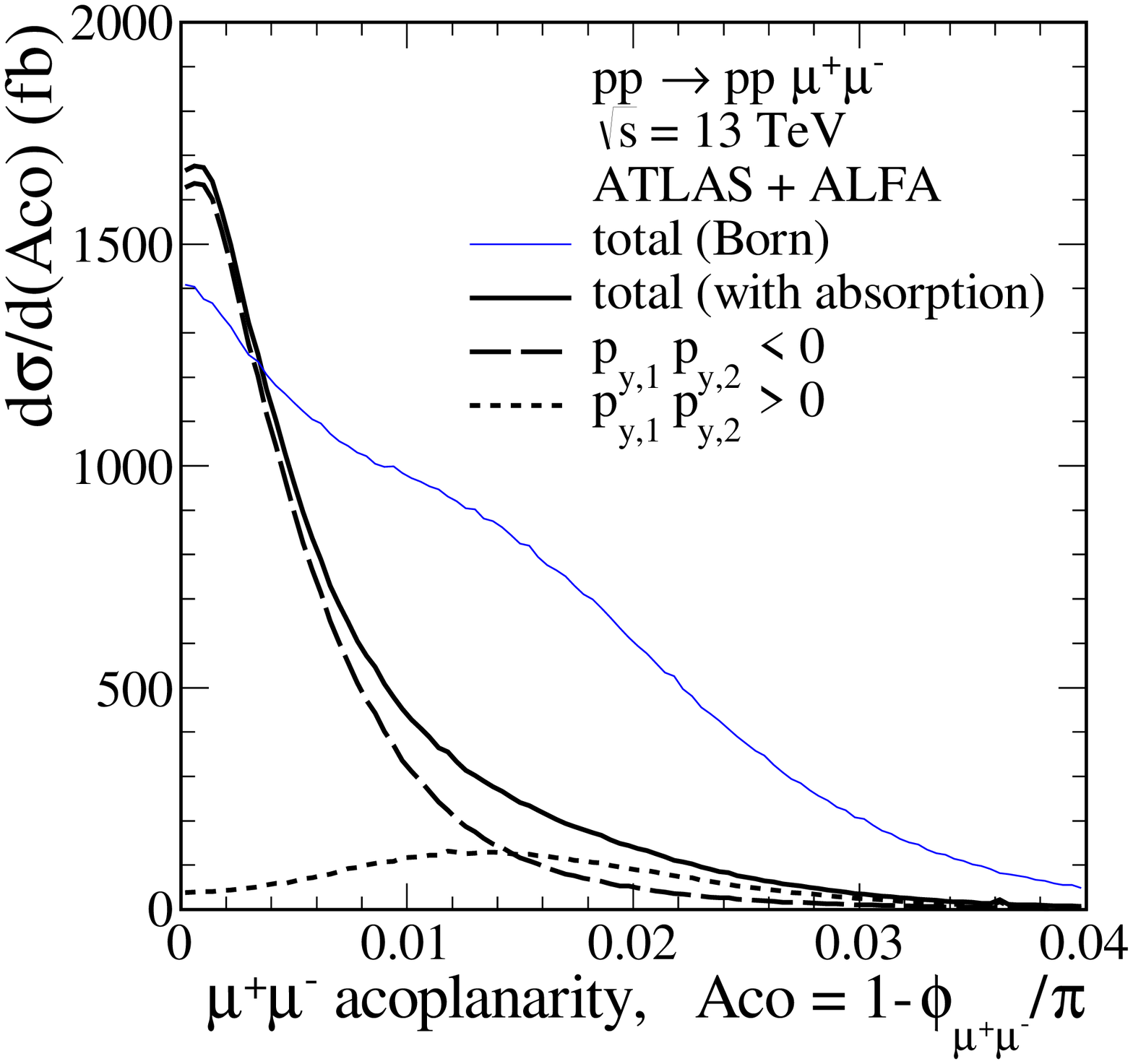}\\
(d)\includegraphics[width=0.46\textwidth]{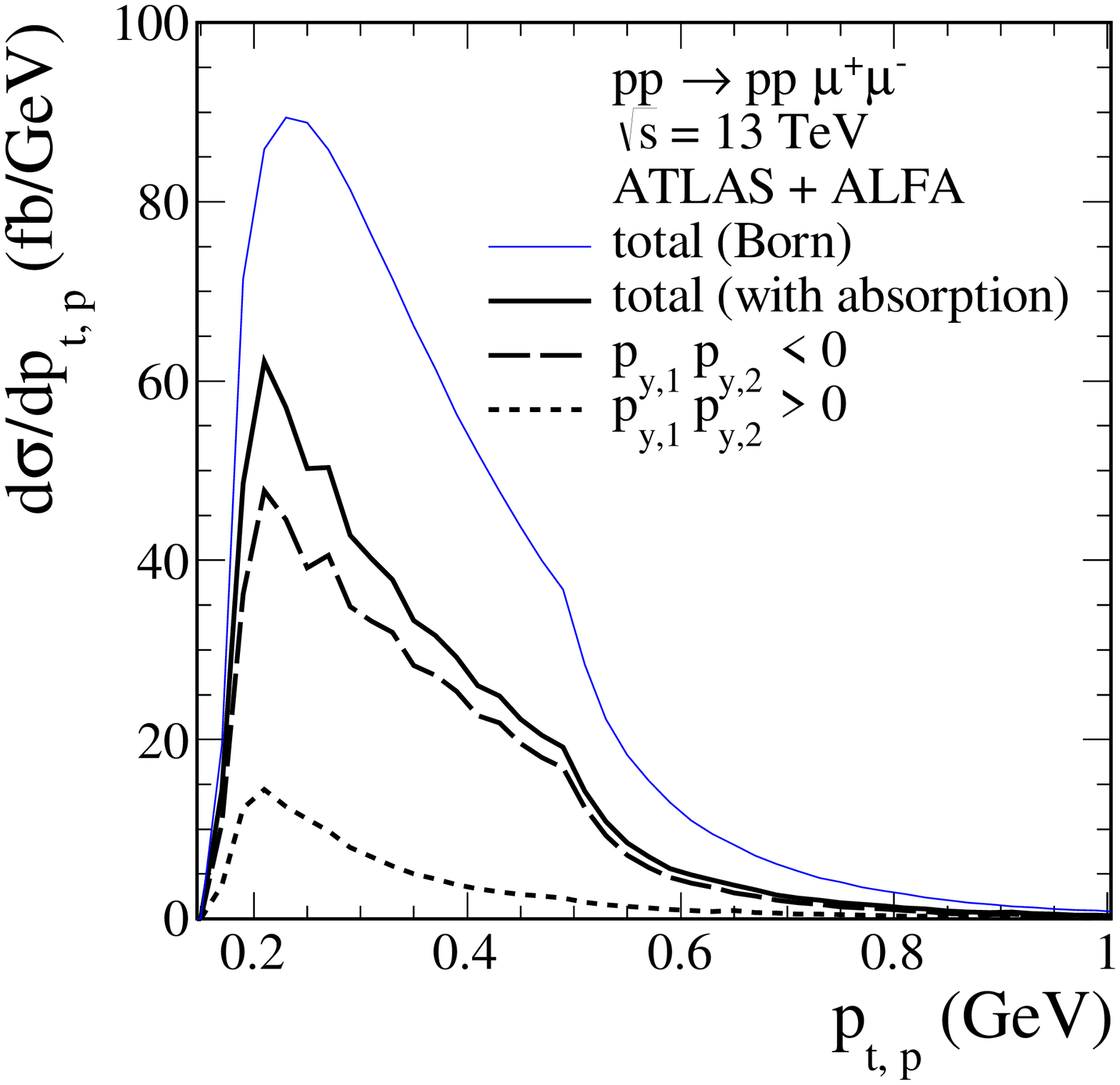}
(e)\includegraphics[width=0.46\textwidth]{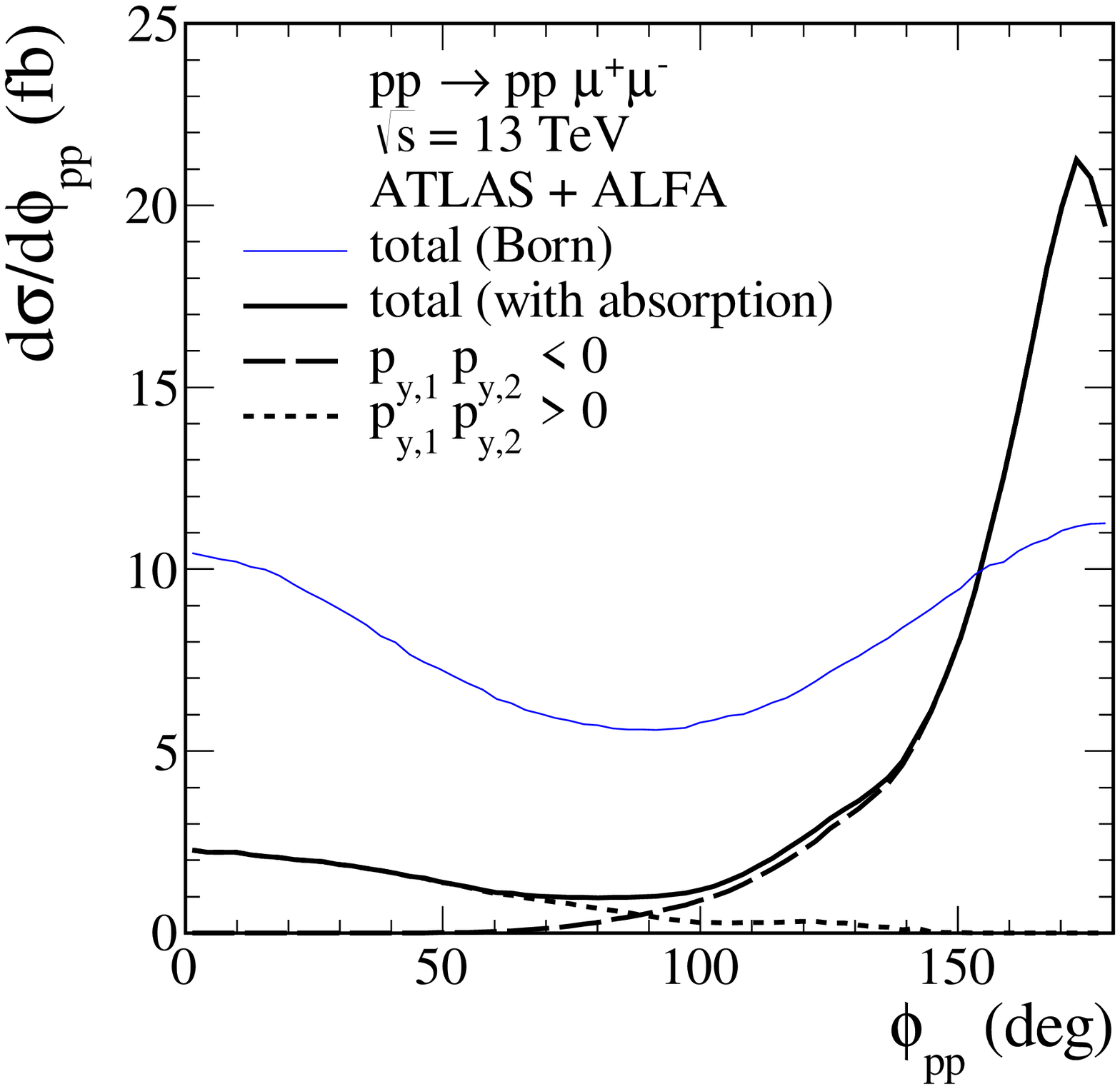}
\caption{\label{fig:ALFA}
\small
The differential cross sections for the $p p \to p p \mu^+ \mu^-$ reactions
for $\sqrt{s} = 13$~TeV and ATLAS + ALFA experimental cuts.
The blue thin lines correspond to the Born results 
while the black thick lines correspond to the results with absorption effects included.
}
\end{figure}
%-----------------------------------------------------------------
In Fig.~\ref{fig:ALFA} we present distributions 
in some observables with the ATLAS + ALFA experimental cuts
without (the blue thin lines) and with (the black thick lines) absorption effects.
Results for two conditions $p_{y,1} p_{y,2} < 0$ (the long-dashed lines) and 
$p_{y,1} p_{y,2} \geqslant 0$ (the dotted lines) and their sum (the solid lines) are shown.
Inclusion of absorption effects modifies the differential distributions
because their shapes depend on the kinematics of outgoing protons.
The measurement of such distributions would allow us to better understand absorption effects.

%--------------------------
\section{Conclusions}
%--------------------------

In the present paper we have explicitly calculated contribution
of the $p p \to p \Delta \mu^+ \mu^-$ and 
$p p \to \Delta \Delta \mu^+ \mu^-$ processes both in momentum space EPA
(using associated photon fluxes derived earlier) and in exact 
kinematically $2 \to 4$ calculation. 
We have considered similar contributions for the Roper resonance ($N(1440)$).
For comparison we have shown also results of calculation obtained
within the $k_t$-factorization approach. 
Using some parametrizations from the literature of the proton structure 
functions one can include also contributions with proton resonances 
in the final state.
Also contributions with single and double continuum dissociation
have been obtained in this way and have been shown for comparison. 

The resonance contributions constitute about 15\%
of the conventional $p p \to p p \mu^+ \mu^-$ cross section 
and leads to an enhancement over the measured recently cross sections 
when ignoring absorption effects.
The resulting cross sections from the $k_t$-factorization approach 
are somewhat larger as those obtained in the explicit $2 \to 4$ calculation.

The $2 \to 4$ calculation allows to include absorption effects on the amplitude level. 
The absorption effects lead to a damping of the cross section. 
The effect depends on the collision energy and kinematical variables.
The corresponding results have been quantified.
However, we have checked numerically that the effect of absorption
for the contributions with one or two $N(1440)$ resonance is larger than
for the conventional $p p \to p p \mu^+ \mu^-$ one.
Our calculations suggest similar effect for the processes
with $\Delta$ resonance production.
But even for the dominant $p p \to p p \mu^+ \mu^-$ process
the resulting cross section overestimate the ATLAS experimental data.

We have shown that in the final comparison with experimental data
one should also take into account contributions when one or both protons dissociate into continuum. 
Naive adding of such contributions would lead to clear overestimation 
of the measured cross sections.
Rapidity gap survival factor associated with remnant fragmentation
seems highly insufficient, see Ref.~\cite{Forthomme:2018sxa}.
%Probably some new mechanisms of absorption effects stay undiscovered.
Clearly some absorption effects are missing.
Multi-parton interactions (see, e.g., Ref.~\cite{Babiarz:2017jxc}) 
are obvious candidates but it is not clear how to include them in a consistent manner.
The parameters of the multi-parton interactions 
are usually adjusted to processes initiated by two gluons but not two photons,
so cannot be directly used in our case.
In our opinion the new experimental data should trigger further studies.

To meet the expectations of the experimental measurement with dedicated forward detectors
we have estimated the cross section for the ATLAS and ALFA experimental cuts.
The cross section for the purely exclusive $pp \to pp \mu^{+} \mu^{-}$ 
reaction, taking into account absorption effects,
is of order of 0.01 pb (\ref{cs_exact_abs_ALFA}).
Several differential distributions have been presented.
We have shown only results for purely non-resonant exclusive component.
What is the role of other semiexclusive processes considered here will
require further experimental and theoretical studies.

%--------------------
\acknowledgments
%--------------------

We are indebted to Leszek Adamczyk, Wolfgang Sch\"afer and 
Rafa{\l} Staszewski for useful discussions.
This research was partially supported by the Polish National Science Centre 
Grant No. 2014/15/B/ST2/02528 and by the Centre for Innovation and
Transfer of Natural Sciences and Engineering Knowledge in Rzesz\'ow.

%------------------------------------------------------------------
\bibliography{refs}

\providecommand{\href}[2]{#2}\begingroup\raggedright\begin{thebibliography}{10}

\bibitem{Aaboud:2017oiq}
M.~Aaboud {\em et~al.}, (ATLAS Collaboration), {\em {Measurement of the
  exclusive $\gamma \gamma \rightarrow \mu^+ \mu^-$ process in proton-proton
  collisions at $\sqrt{s} = 13$ TeV with the ATLAS detector},}
  \href{http://dx.doi.org/10.1016/j.physletb.2017.12.043}{Phys. Lett.
  {\bfseries B777} (2018) 303},
\href{http://arxiv.org/abs/1708.04053}{{arXiv:1708.04053 [hep-ex]}}.
%%CITATION = ARXIV:1708.04053;%%.

\bibitem{Chatrchyan:2011ci}
S.~Chatrchyan {\em et~al.}, (CMS Collaboration), {\em {Exclusive $\gamma \gamma
  \rightarrow \mu^+ \mu^-$ production in proton-proton collisions at
  $\sqrt{s}=7$ TeV},} \href{http://dx.doi.org/10.1007/JHEP01(2012)052}{JHEP
  {\bfseries 01} (2012) 052},
\href{http://arxiv.org/abs/1111.5536}{{arXiv:1111.5536 [hep-ex]}}.
%%CITATION = ARXIV:1111.5536;%%.

\bibitem{Chatrchyan:2012tv}
S.~Chatrchyan {\em et~al.}, (CMS Collaboration), {\em {Search for exclusive or
  semi-exclusive $\gamma \gamma$ production and observation of exclusive and
  semi-exclusive $e^{+} e^{-}$ production in $pp$ collisions at $\sqrt{s}=7$
  TeV},} \href{http://dx.doi.org/10.1007/JHEP11(2012)080}{JHEP {\bfseries 11}
  (2012) 080},
\href{http://arxiv.org/abs/1209.1666}{{arXiv:1209.1666 [hep-ex]}}.
%%CITATION = ARXIV:1209.1666;%%.

\bibitem{Aad:2015bwa}
G.~Aad {\em et~al.}, (ATLAS Collaboration), {\em {Measurement of exclusive
  $\gamma \gamma \rightarrow \ell^+\ell^-$ production in proton-proton
  collisions at $\sqrt{s} = 7$ TeV with the ATLAS detector},}
  \href{http://dx.doi.org/10.1016/j.physletb.2015.07.069}{Phys. Lett.
  {\bfseries B749} (2015) 242},
\href{http://arxiv.org/abs/1506.07098}{{arXiv:1506.07098 [hep-ex]}}.
%%CITATION = ARXIV:1506.07098;%%.

\bibitem{Cms:2018het}
A.~M. Sirunyan {\em et~al.}, (CMS and TOTEM Collaborations), {\em {Observation
  of proton-tagged, central (semi)exclusive production of high-mass lepton
  pairs in $pp$ collisions at 13 TeV with the CMS-TOTEM precision proton
  spectrometer},} CMS-PPS-17-001, TOTEM-2018-001, CERN-EP-2018-014, Submitted
  to: JHEP (2018) ,
\href{http://arxiv.org/abs/1803.04496}{{arXiv:1803.04496 [hep-ex]}}.
%%CITATION = ARXIV:1803.04496;%%.

\bibitem{Lebiedowicz:2012gg}
P.~Lebiedowicz, R.~Pasechnik, and A.~Szczurek, {\em {QCD diffractive mechanism
  of exclusive $W^{+} W^{-}$ pair production at high energies},} Nucl. Phys.
  {\bfseries B867} (2013) 61,
\href{http://arxiv.org/abs/1203.1832}{{arXiv:1203.1832 [hep-ph]}}.
%\%CITATION = ARXIV:1203.1832;\%\%.

\bibitem{Lebiedowicz:2015cea}
P.~Lebiedowicz and A.~Szczurek, {\em {Exclusive production of heavy charged
  Higgs boson pairs in the $p p \to p p H^+ H^-$ reaction at the LHC and a
  future circular collider},}
  \href{http://dx.doi.org/10.1103/PhysRevD.91.095008}{Phys. Rev. {\bfseries
  D91} (2015) 095008},
\href{http://arxiv.org/abs/1502.03323}{{arXiv:1502.03323 [hep-ph]}}.
%%CITATION = ARXIV:1502.03323;%%.

\bibitem{Lebiedowicz:2016lmn}
P.~Lebiedowicz, M.~Luszczak, R.~Pasechnik, and A.~Szczurek, {\em {Can the
  diphoton enhancement at 750 GeV be due to a neutral technipion?},}
  \href{http://dx.doi.org/10.1103/PhysRevD.94.015023}{Phys. Rev. {\bfseries
  D94} no.~1, (2016) 015023},
\href{http://arxiv.org/abs/1604.02037}{{arXiv:1604.02037 [hep-ph]}}.
%%CITATION = ARXIV:1604.02037;%%.

\bibitem{Dyndal:2014yea}
M.~Dyndal and L.~Schoeffel, {\em {The role of finite-size effects on the
  spectrum of equivalent photons in proton-proton collisions at the LHC},}
  \href{http://dx.doi.org/10.1016/j.physletb.2014.12.019}{Phys. Lett.
  {\bfseries B741} (2015) 66},
\href{http://arxiv.org/abs/1410.2983}{{arXiv:1410.2983 [hep-ph]}}.
%%CITATION = ARXIV:1410.2983;%%.

\bibitem{Schafer:2007mm}
W.~Sch{\"a}fer and A.~Szczurek, {\em {Exclusive photoproduction of $J/\psi$ in
  proton-proton and proton-antiproton scattering},}
  \href{http://dx.doi.org/10.1103/PhysRevD.76.094014}{Phys.Rev. {\bfseries D76}
  (2007) 094014},
\href{http://arxiv.org/abs/0705.2887}{{arXiv:0705.2887 [hep-ph]}}.
%\%CITATION = ARXIV:0705.2887;\%\%.

\bibitem{Lebiedowicz:2009pj}
P.~Lebiedowicz and A.~Szczurek, {\em {Exclusive $pp \to pp \pi^{+}\pi^{-}$
  reaction: From the threshold to LHC},}
  \href{http://dx.doi.org/10.1103/PhysRevD.81.036003}{Phys.Rev. {\bfseries D81}
  (2010) 036003},
\href{http://arxiv.org/abs/0912.0190}{{arXiv:0912.0190 [hep-ph]}}.
%\%CITATION = ARXIV:0912.0190;\%\%.

\bibitem{Lebiedowicz:2015eka}
P.~Lebiedowicz and A.~Szczurek, {\em {Revised model of absorption corrections
  for the $p p \to p p \pi^{+} \pi^{-}$ process},}
  \href{http://dx.doi.org/10.1103/PhysRevD.92.054001}{Phys. Rev. {\bfseries
  D92} (2015) 054001},
\href{http://arxiv.org/abs/1504.07560}{{arXiv:1504.07560 [hep-ph]}}.
%%CITATION = ARXIV:1504.07560;%%.

\bibitem{Lebiedowicz:2016ioh}
P.~Lebiedowicz, O.~Nachtmann, and A.~Szczurek, {\em {Central exclusive
  diffractive production of the $\pi^{+}\pi^{-}$ continuum, scalar, and tensor
  resonances in $pp$ and $p \bar{p}$ scattering within the tensor Pomeron
  approach},} \href{http://dx.doi.org/10.1103/PhysRevD.93.054015}{Phys. Rev.
  {\bfseries D93} (2016) 054015},
\href{http://arxiv.org/abs/1601.04537}{{arXiv:1601.04537 [hep-ph]}}.
%%CITATION = ARXIV:1601.04537;%%.

\bibitem{Lebiedowicz:2011tp}
P.~Lebiedowicz and A.~Szczurek, {\em {$pp \to pp K^{+}K^{-}$ reaction at high
  energies},} \href{http://dx.doi.org/10.1103/PhysRevD.85.014026}{Phys.Rev.
  {\bfseries D85} (2012) 014026},
\href{http://arxiv.org/abs/1110.4787}{{arXiv:1110.4787 [hep-ph]}}.
%\%CITATION = ARXIV:1110.4787;\%\%.

\bibitem{Lebiedowicz:2018eui}
P.~Lebiedowicz, O.~Nachtmann, and A.~Szczurek, {\em {Towards a complete study
  of central exclusive production of $K^{+}K^{-}$ pairs in proton-proton
  collisions within the tensor Pomeron approach},}
  \href{http://dx.doi.org/10.1103/PhysRevD.98.014001}{Phys. Rev. {\bfseries
  D98} (2018) 014001},
\href{http://arxiv.org/abs/1804.04706}{{arXiv:1804.04706 [hep-ph]}}.
%%CITATION = ARXIV:1804.04706;%%.

\bibitem{Lebiedowicz:2018sdt}
P.~Lebiedowicz, O.~Nachtmann, and A.~Szczurek, {\em {Central exclusive
  diffractive production of $p \bar{p}$ pairs in proton-proton collisions at
  high energies},} \href{http://dx.doi.org/10.1103/PhysRevD.97.094027}{Phys.
  Rev. {\bfseries D97} no.~9, (2018) 094027},
\href{http://arxiv.org/abs/1801.03902}{{arXiv:1801.03902 [hep-ph]}}.
%%CITATION = ARXIV:1801.03902;%%.

\bibitem{Baur:1998ay}
G.~Baur, K.~Hencken, and D.~Trautmann, {\em {Photon-photon physics in very
  peripheral collisions of relativistic heavy ions},}
  \href{http://dx.doi.org/10.1088/0954-3899/24/9/003}{J. Phys. {\bfseries G24}
  (1998) 1657},
\href{http://arxiv.org/abs/hep-ph/9804348}{{arXiv:hep-ph/9804348 [hep-ph]}}.
%%CITATION = HEP-PH/9804348;%%.

\bibitem{Guzey:2014axa}
V.~Guzey and M.~Zhalov, {\em {Photoproduction of $J/\psi$ and $\psi(2S)$ in
  proton-proton ultraperipheral collisions at the LHC},}
\href{http://arxiv.org/abs/1405.7529}{{arXiv:1405.7529 [hep-ph]}}.
%%CITATION = ARXIV:1405.7529;%%.

\bibitem{Cisek:2016kvr}
A.~Cisek, W.~Sch{\"a}fer, and A.~Szczurek, {\em {Semiexclusive production of
  $J/\psi$ mesons in proton-proton collisions with electromagnetic and
  diffractive dissociation of one of the protons},}
  \href{http://dx.doi.org/10.1016/j.physletb.2017.03.048}{Phys. Lett.
  {\bfseries B769} (2017) 176},
\href{http://arxiv.org/abs/1611.08210}{{arXiv:1611.08210 [hep-ph]}}.
%%CITATION = ARXIV:1611.08210;%%.

\bibitem{daSilveira:2014jla}
G.~G. da~Silveira, L.~Forthomme, K.~Piotrzkowski, W.~Sch{\"a}fer, and
  A.~Szczurek, {\em {Central $\mu^+ \mu^-$ production via photon-photon fusion
  in proton-proton collisions with proton dissociation},}
  \href{http://dx.doi.org/10.1007/JHEP02(2015)159}{JHEP {\bfseries 02} (2015)
  159},
\href{http://arxiv.org/abs/1409.1541}{{arXiv:1409.1541 [hep-ph]}}.
%%CITATION = ARXIV:1409.1541;%%.

\bibitem{Luszczak:2015aoa}
M.~{\L}uszczak, W.~Sch{\"a}fer, and A.~Szczurek, {\em {Two-photon dilepton
  production in proton-proton collisions: Two alternative approaches},}
  \href{http://dx.doi.org/10.1103/PhysRevD.93.074018}{Phys. Rev. {\bfseries
  D93} no.~7, (2016) 074018},
\href{http://arxiv.org/abs/1510.00294}{{arXiv:1510.00294 [hep-ph]}}.
%%CITATION = ARXIV:1510.00294;%%.

\bibitem{daSilveira:2015hha}
G.~G. da~Silveira and V.~P. Goncalves, {\em {Constraining the photon flux in
  two-photon processes at the LHC},}
  \href{http://dx.doi.org/10.1103/PhysRevD.92.014013}{Phys. Rev. {\bfseries
  D92} no.~1, (2015) 014013},
\href{http://arxiv.org/abs/1506.01352}{{arXiv:1506.01352 [hep-ph]}}.
%%CITATION = ARXIV:1506.01352;%%.

\bibitem{Goncalves:2018gca}
V.~P. Goncalves, M.~M. Jaime, D.~E. Martins, and M.~S. Rangel, {\em {Exclusive
  and diffractive $\mu^+ \mu^-$ production in $pp$ collisions at the LHC},}
  \href{http://dx.doi.org/10.1103/PhysRevD.97.074024}{Phys. Rev. {\bfseries
  D97} no.~7, (2018) 074024},
\href{http://arxiv.org/abs/1802.07339}{{arXiv:1802.07339 [hep-ph]}}.
%%CITATION = ARXIV:1802.07339;%%.

\bibitem{Aznauryan:2009mx}
I.~G. Aznauryan {\em et~al.}, (CLAS Collaboration), {\em {Electroexcitation of
  nucleon resonances from CLAS data on single pion electroproduction},}
  \href{http://dx.doi.org/10.1103/PhysRevC.80.055203}{Phys. Rev. {\bfseries
  C80} (2009) 055203},
\href{http://arxiv.org/abs/0909.2349}{{arXiv:0909.2349 [nucl-ex]}}.
%%CITATION = ARXIV:0909.2349;%%.

\bibitem{Mokeev:2015lda}
V.~I. Mokeev {\em et~al.}, {\em {New results from the studies of the
  $N(1440)1/2^+$, $N(1520)3/2^-$, and $\Delta(1620)1/2^-$ resonances in
  exclusive $ep \to e'p' \pi^+ \pi^-$ electroproduction with the CLAS
  detector},} \href{http://dx.doi.org/10.1103/PhysRevC.93.025206}{Phys. Rev.
  {\bfseries C93} no.~2, (2016) 025206},
\href{http://arxiv.org/abs/1509.05460}{{arXiv:1509.05460 [nucl-ex]}}.
%%CITATION = ARXIV:1509.05460;%%.

\bibitem{Tiator:2008kd}
L.~Tiator and M.~Vanderhaeghen, {\em {Empirical transverse charge densities in
  the nucleon-to-$P_{11}(1440)$ transition},}
  \href{http://dx.doi.org/10.1016/j.physletb.2009.01.048}{Phys. Lett.
  {\bfseries B672} (2009) 344},
\href{http://arxiv.org/abs/0811.2285}{{arXiv:0811.2285 [hep-ph]}}.
%%CITATION = ARXIV:0811.2285;%%.

\bibitem{Ramalho:2017muv}
G.~Ramalho, {\em {Analytic parametrizations of the $\gamma^{*} N \to N(1440)$
  form factors inspired by light-front holography},}
  \href{http://dx.doi.org/10.1103/PhysRevD.96.054021}{Phys. Rev. {\bfseries
  D96} no.~5, (2017) 054021},
\href{http://arxiv.org/abs/1706.05707}{{arXiv:1706.05707 [hep-ph]}}.
%%CITATION = ARXIV:1706.05707;%%.

\bibitem{Ramalho:2014hia}
G.~Ramalho and K.~Tsushima, {\em {$\gamma^{*} N \to N(1710)$ transition at high
  momentum transfer},}
  \href{http://dx.doi.org/10.1103/PhysRevD.89.073010}{Phys. Rev. {\bfseries
  D89} no.~7, (2014) 073010},
\href{http://arxiv.org/abs/1402.3234}{{arXiv:1402.3234 [hep-ph]}}.
%%CITATION = ARXIV:1402.3234;%%.

\bibitem{Gutsche:2017lyu}
T.~Gutsche, V.~E. Lyubovitskij, and I.~Schmidt, {\em {Electromagnetic structure
  of nucleon and Roper in soft-wall AdS/QCD},}
  \href{http://dx.doi.org/10.1103/PhysRevD.97.054011}{Phys. Rev. {\bfseries
  D97} no.~5, (2018) 054011},
\href{http://arxiv.org/abs/1712.08410}{{arXiv:1712.08410 [hep-ph]}}.
%%CITATION = ARXIV:1712.08410;%%.

\bibitem{Ramalho:2017pyc}
G.~Ramalho and D.~Melnikov, {\em {Valence quark contributions for the
  $\gamma^\ast N \to N(1440)$ form factors from light-front holography},}
  \href{http://dx.doi.org/10.1103/PhysRevD.97.034037}{Phys. Rev. {\bfseries
  D97} no.~3, (2018) 034037},
\href{http://arxiv.org/abs/1703.03819}{{arXiv:1703.03819 [hep-ph]}}.
%%CITATION = ARXIV:1703.03819;%%.

\bibitem{Jones:1972ky}
H.~F. Jones and M.~D. Scadron, {\em {Multipole $\gamma N- \Delta$ Form Factors
  and Resonant Photo- and Electroproduction},}
\href{http://dx.doi.org/10.1016/0003-4916(73)90476-4}{Annals Phys. {\bfseries
  81} (1973) 1}.
%%CITATION = APNYA,81,1;%%.

\bibitem{Gail:2005gz}
T.~A. Gail and T.~R. Hemmert, {\em {Signatures of chiral dynamics in the
  nucleon to Delta transition},}
  \href{http://dx.doi.org/10.1140/epja/i2006-10023-y}{Eur. Phys. J. {\bfseries
  A28} (2006) 91},
\href{http://arxiv.org/abs/nucl-th/0512082}{{arXiv:nucl-th/0512082 [nucl-th]}}.
%%CITATION = NUCL-TH/0512082;%%.

\bibitem{Ramalho:2008dp}
G.~Ramalho, M.~T. Pena, and F.~Gross, {\em {D-state effects in the
  electromagnetic $N \Delta$ transition},}
  \href{http://dx.doi.org/10.1103/PhysRevD.78.114017}{Phys. Rev. {\bfseries
  D78} (2008) 114017},
\href{http://arxiv.org/abs/0810.4126}{{arXiv:0810.4126 [hep-ph]}}.
%%CITATION = ARXIV:0810.4126;%%.

\bibitem{Budnev:1974de}
V.~M. Budnev, I.~F. Ginzburg, G.~V. Meledin, and V.~G. Serbo, {\em {The
  two-photon particle production mechanism. Physical problems. Applications.
  Equivalent photon approximation},}
\href{http://dx.doi.org/10.1016/0370-1573(75)90009-5}{Phys. Rept. {\bfseries
  15} (1975) 181}.
%%CITATION = PRPLC,15,181;%%.

\bibitem{Fiore:2002re}
R.~Fiore, A.~Flachi, L.~L. Jenkovszky, A.~I. Lengyel, and V.~K. Magas, {\em
  {Explicit model realizing parton-hadron duality},}
  \href{http://dx.doi.org/10.1140/epja/i2002-10047-3}{Eur. Phys. J. {\bfseries
  A15} (2002) 505},
\href{http://arxiv.org/abs/hep-ph/0206027}{{arXiv:hep-ph/0206027 [hep-ph]}}.
%%CITATION = HEP-PH/0206027;%%.

\bibitem{Fiore:2003dg}
R.~Fiore, A.~Flachi, L.~L. Jenkovszky, A.~I. Lengyel, and V.~K. Magas, {\em
  {Kinematically complete analysis of the CLAS data on the proton structure
  function $F_{2}$ in a Regge-dual model},}
  \href{http://dx.doi.org/10.1103/PhysRevD.69.014004}{Phys. Rev. {\bfseries
  D69} (2004) 014004},
\href{http://arxiv.org/abs/hep-ph/0308178}{{arXiv:hep-ph/0308178 [hep-ph]}}.
%%CITATION = HEP-PH/0308178;%%.

\bibitem{Szczurek:1999rd}
A.~Szczurek and V.~Uleshchenko, {\em {Nonpartonic components in the nucleon
  structure functions at small $Q^2$ in the broad range of $x$},}
  \href{http://dx.doi.org/10.1007/s100520000218}{Eur. Phys. J. {\bfseries C12}
  (2000) 663},
\href{http://arxiv.org/abs/hep-ph/9904288}{{arXiv:hep-ph/9904288 [hep-ph]}}.
%%CITATION = HEP-PH/9904288;%%.

\bibitem{Forthomme:2018sxa}
L.~Forthomme, M.~{\L}uszczak, W.~Sch{\"a}fer, and A.~Szczurek, {\em {Rapidity
  gap survival factors caused by remnant fragmentation for $W^+ W^-$ pair
  production via $\gamma^* \gamma^* \to W^+ W^-$ subprocess with photon
  transverse momenta},}
\href{http://arxiv.org/abs/1805.07124}{{arXiv:1805.07124 [hep-ph]}}.
%%CITATION = ARXIV:1805.07124;%%.

\bibitem{Harland-Lang:2015cta}
L.~A. Harland-Lang, V.~A. Khoze, and M.~G. Ryskin, {\em {Exclusive physics at
  the LHC with SuperChic 2},}
  \href{http://dx.doi.org/10.1140/epjc/s10052-015-3832-8}{Eur. Phys. J.
  {\bfseries C76} no.~1, (2016) 9},
\href{http://arxiv.org/abs/1508.02718}{{arXiv:1508.02718 [hep-ph]}}.
%%CITATION = ARXIV:1508.02718;%%.

\bibitem{Babiarz:2017jxc}
I.~Babiarz, R.~Staszewski, and A.~Szczurek, {\em {Multi-parton interactions and
  rapidity gap survival probability in jet-gap-jet processes},}
  \href{http://dx.doi.org/10.1016/j.physletb.2017.05.095}{Phys. Lett.
  {\bfseries B771} (2017) 532},
\href{http://arxiv.org/abs/1704.00546}{{arXiv:1704.00546 [hep-ph]}}.
%%CITATION = ARXIV:1704.00546;%%.

\bibitem{Khoze:2017sdd}
V.~A. Khoze, A.~D. Martin, and M.~G. Ryskin, {\em {Multiple interactions and
  rapidity gap survival},} \href{http://dx.doi.org/10.1088/1361-6471/aab1bf}{J.
  Phys. {\bfseries G45} no.~5, (2018) 053002},
\href{http://arxiv.org/abs/1710.11505}{{arXiv:1710.11505 [hep-ph]}}.
%%CITATION = ARXIV:1710.11505;%%.

\bibitem{Adamczyk_priv}
L.~Adamczyk.
\newblock {Private communication}.

\end{thebibliography}\endgroup
%------------------------------------------------------------------

\end{document}